\documentclass[10pt, conference, letterpaper]{IEEEtran}

\usepackage{cite}
\usepackage{amsmath,amssymb,amsfonts}
\usepackage{algorithmic}
\usepackage{graphicx}
\usepackage{textcomp}
\usepackage{xcolor}
\def\BibTeX{{\rm B\kern-.05em{\sc i\kern-.025em b}\kern-.08em
    T\kern-.1667em\lower.7ex\hbox{E}\kern-.125emX}}

\usepackage{booktabs}
\usepackage{color, colortbl}
\usepackage{tabularx}
\usepackage{soul}
\usepackage{balance}

\usepackage{graphicx}
\usepackage{subfig}
\usepackage{blindtext}
\usepackage{adjustbox}
\usepackage{multirow}
\usepackage{color}
\usepackage{booktabs}
\usepackage{tabularx}
\usepackage{colortbl}

\usepackage{bbding}

\usepackage{tikz}
\usepackage{listings}
\usepackage{etoolbox}
\usepackage{subfig}
\usepackage{url}
\usepackage{setspace}
\usepackage[english]{babel}
\usepackage{color, colortbl}
\usepackage{enumitem}
\usepackage{balance}
\usepackage{tabularx}

\definecolor{Gray}{gray}{0.9}

\newcommand{\circled}[2][]{\tikz[baseline=(char.base)]
	{\node[shape = circle, draw, inner sep = 1pt]
		(char) {\phantom{\ifblank{#1}{#2}{#1}}};%
		\node at (char.center) {\makebox[0pt][c]{#2}};}}
\robustify{\circled}

\newcommand{\Interactive}{\texttt{interactive}\xspace}

\newcommand{\SystemName}{\textsc{Camel}\xspace}

\makeatletter
\def\@IEEEsectpunct{.\ \,}
\def\paragraph{\@startsection{paragraph}{4}{\z@}{1.5ex plus 1.5ex minus 0.5ex}%
{0ex}{\normalfont\normalsize\sffamily\bfseries}}

\newcommand\cparagraph[1]{\vspace{2mm} \noindent \textbf{#1}\xspace}

\DeclareMathOperator*{\argmin}{arg\,min}

\begin{document}
\title{Smart, Adaptive Energy Optimization for Mobile Web Interactions}

\author{}

\author{\IEEEauthorblockN{Jie Ren\IEEEauthorrefmark{2}, Lu Yuan\IEEEauthorrefmark{3}, Petteri Nurmi\IEEEauthorrefmark{6}, Xiaoming Wang\IEEEauthorrefmark{2}, Miao Ma\IEEEauthorrefmark{2},
Ling Gao\IEEEauthorrefmark{3}, \\Zhanyong Tang\IEEEauthorrefmark{3}, Jie Zheng\IEEEauthorrefmark{3}, Zheng
Wang\IEEEauthorrefmark{1}}

\IEEEauthorblockA{
\IEEEauthorrefmark{2}Shannxi Normal University, China,
\IEEEauthorrefmark{6}University of Helsinki, Finland, \\
\IEEEauthorrefmark{3}Northwest University, China,
\IEEEauthorrefmark{1}University of Leeds, UK \\
} }


\maketitle

\begin{abstract}
Web technology underpins many interactive mobile applications. However, energy-efficient mobile web interactions is an outstanding
challenge. Given the increasing diversity and complexity of mobile hardware, any practical optimization scheme must work for a wide range
of users, mobile platforms and web workloads. This paper presents \SystemName, a novel energy optimization system for mobile web
interactions. \SystemName leverages machine learning techniques to develop a smart, adaptive scheme to judiciously trade performance for
reduced power consumption. Unlike prior work, \SystemName directly models how a given web content affects the user expectation and uses
this to guide energy optimization. It goes further by employing transfer learning and conformal predictions to tune a previously learned
model in the end-user environment and improve it over time. We apply \SystemName to Chromium and evaluate it on four distinct mobile
systems involving 1,000 testing webpages and 30 users. Compared to four state-of-the-art web-event optimizers, \SystemName delivers 22\%
more energy savings, but with 49\% fewer violations on the quality of user experience, and exhibits orders of magnitudes less overhead
when targeting a new computing environment.
\end{abstract}


\section{Introduction}
 Web has become the main approach for accessing information on mobile systems. Indeed, recent studies suggest that 70\% of
all web traffic comes from mobile devices with the average mobile user in the US spending over three hours per day with web
content~\cite{mobilestudy}. When interacting with web content, mobile users want their devices to react fast to interaction events while
having a long-lasting battery~\cite{Zuniga:2019:THQ:3308558.3313428}. Achieving both at once is difficult as web content access often comes
with a high energy cost to end-users~\cite{thiagarajan2012killed} but existing mechanisms for web access optimization often ignore the
effects of energy savings on user experience~\cite{ren2017optimise,li2016automated,zhu2015event,Peters:2018:PWB:3205289.3205293}.

Prior work on energy optimization for mobile web access has predominantly focused on lowering power consumption of the transmission and
rendering operations for loading a web page~\cite{ren2017optimise,Ren:2018:PNW:3281411.3281422,8366936, 6848020,li2016automated}.
Unfortunately, these approaches can only achieve modest savings as they ignore the continuous nature of web interactions.
Due to small-form-factor of mobile devices, the webpages often can only be seen through multiple user interactions,
such as scrolling and zooming. As we will show in this paper, these operations can consume 2 to 5 times more energy than the initial page
loading phase and hence optimizing energy drain of these operations is critical.

Some more recent works try to reduce the energy footprint of web interactions by limiting the processor clock speed~\cite{zhu2015event,
Peters:2018:PWB:3205289.3205293, ml2019} or dropping some interaction events~\cite{xu2018ebrowser}. However, these solutions are suboptimal
as they achieve energy savings at the cost of user experience. Indeed, the user's sensitivity to response delay differs,  with the content
type, nature of interactions and interactive speed all affecting user expectations~\cite{seeker2014measuring}. Another drawback of all
existing approaches is that they offer little ability to adapt a decision policy across different computing environments. As mobile
hardware, web workloads, and operating system internal mechanisms change over time, it is unlikely that a policy developed today will
remain suitable for tomorrow.

We present \SystemName, a novel energy optimization strategy for mobile web interactions that takes into consideration both the need to
reduce the energy footprint and to provide good user experience. \SystemName preserves user experience through machine learning
models learned \textit{offline} and deployed on the device. These models capture subtle interactions between web content and user perception of delay. This enables \SystemName to make
energy-efficient scheduling decisions for any \emph{new} webpage \emph{unseen at design time}. Specifically, \SystemName integrates two
types of machine-learning models: a per-user specific predictor that estimates the minimum acceptable response delay for given web
content, and a profit estimator that assesses the outcome of different scheduling decisions given the expected user interaction
requirements. \SystemName uses these two predictors to quickly find the optimal processing configuration that consumes the least energy but
still meeting the interactivity target of the user.

Developing a practical machine learning approach that can generalize across a diverse range of constantly evolving hardware architectures,
web workloads and user habits is far from trivial. Prior work has addressed this portability issue through rewriting or
retraining~\cite{mlcpieee}. Both solutions, however, are inadequate for mobile web optimization as they either require expert involvement
or substantial overhead for training data collection. \SystemName is designed to avoid this pitfall.

To target a diverse set of users and mobile devices, \SystemName employs a novel transfer learning~\cite{long2017deep} based approach to
tune a  generic model developed by the application vendor to match the requirements of a new user or hardware platform. Our insight is that
one can re-use the knowledge previously obtained from a different platform or user to speed up learning in a new environment considerably.
Instead of gathering training samples by profiling the entire dataset, we use statistical methods to determine which part of
the dataset is likely to offer useful representative information. By using fewer training instances, we reduce the profiling times and
end-user involvement as well as the cost associated with them. We show that despite using many fewer training instances, the resultant
performance of \SystemName is comparable to retraining from scratch by profiling the entire dataset.

To adapt to changes in the deployment environment, \SystemName combines statistical and probabilistic assessments to estimate the error
bound (or credibility) of a prediction. This provides a rigorous methodology to quantify how much we should trust a model's output,
allowing a learning framework to use feedback on uncertain inputs to continuously update a decision model in the end-user environment.

We demonstrate the benefits of \SystemName by integrating it into the rendering architecture of Chromium~\cite{chromium} and evaluating it
against four event-based web schedulers~\cite{ml2019,xu2018ebrowser,phaseaware,zhu2015event}. We perform an unprecedentedly large-scale
evaluation involving 1,000 testing webpages, 30 users and four distinct mobile devices. Experimental results show that \SystemName
consistently outperforms all existing schemes by delivering better energy savings with less frequent violations on the
quality-of-experience (QoE). We consider the cases for porting an existing model to different users or hardware. We show that \SystemName
provides portable performance but incurs significantly less training overhead over prior strategies.

\cparagraph{Contributions.} This paper is the first to:
\begin{itemize}
\item show how a content-aware QoE optimizing scheme can be developed for web interactions using predictive modeling
    (Section~\ref{sec:pred});
\item employ transfer learning to address the model portability issue across users and devices (Section~\ref {sec:hardwareport});
\item exploit statistical assessments to detect and improve ageing models for mobile web browsing (Section~\ref{sec:agining}).
\end{itemize}

\section{Background and Motivation}


\SystemName reduces energy usage during web interactions. Existing works largely optimize the initial page loading phase, but as we
demonstrate below, interactions have higher energy drain and thus more potential for savings. The few works ~\cite{zhu2015event,
phaseaware,ml2019} to address interactions assume a fixed response deadline for web content, but this runs the risk of degrading the
overall user experience. By contrast, \SystemName minimizes energy consumption without compromising QoE, by offering ``sufficiently good"
performance. This is motivated by user experience studies showing that improvements beyond ``good enough" are not guaranteed to enhance
mobile user experience~\cite{Choi:2019:GPG:3307334.3326075,Zuniga:2019:THQ:3308558.3313428}, e.g., the user cannot tell the difference
between a lag of 10ms compared to a lag of 100ms~\cite{seeker2014measuring}.

\subsection{Problem Scope}
 \SystemName targets interactions taking place after web contents have been fetched and the Document
Object Model (DOM) tree constructed. We consider three representative browsing gestures: scrolling, pinching (i.e.,
zoom in and out), and flinging (a quick swipe). We do not consider clicking because it often leads to a new page
loading which can be optimized by a page-loading-specific scheme like ~\cite{Ren:2018:PNW:3281411.3281422}. Our work
targets the widely used big.LITTLE~\cite{biglittle} mobile architecture. As a case study, we apply \SystemName to
Chromium,  the open-source project behind Google Chrome and many other browsers like Opera and Microsoft Edge for ARM.
Note that as \SystemName targets response to interaction events within the web rendering engine, it is not restricted
to browsers but equally applicable to webview-based apps like social media and newsreaders.

\subsection{Motivating Examples}

\begin{figure}[!t]
	\centering
	\subfloat[][Scrolling up webpages]{\includegraphics[width=0.25\textwidth]{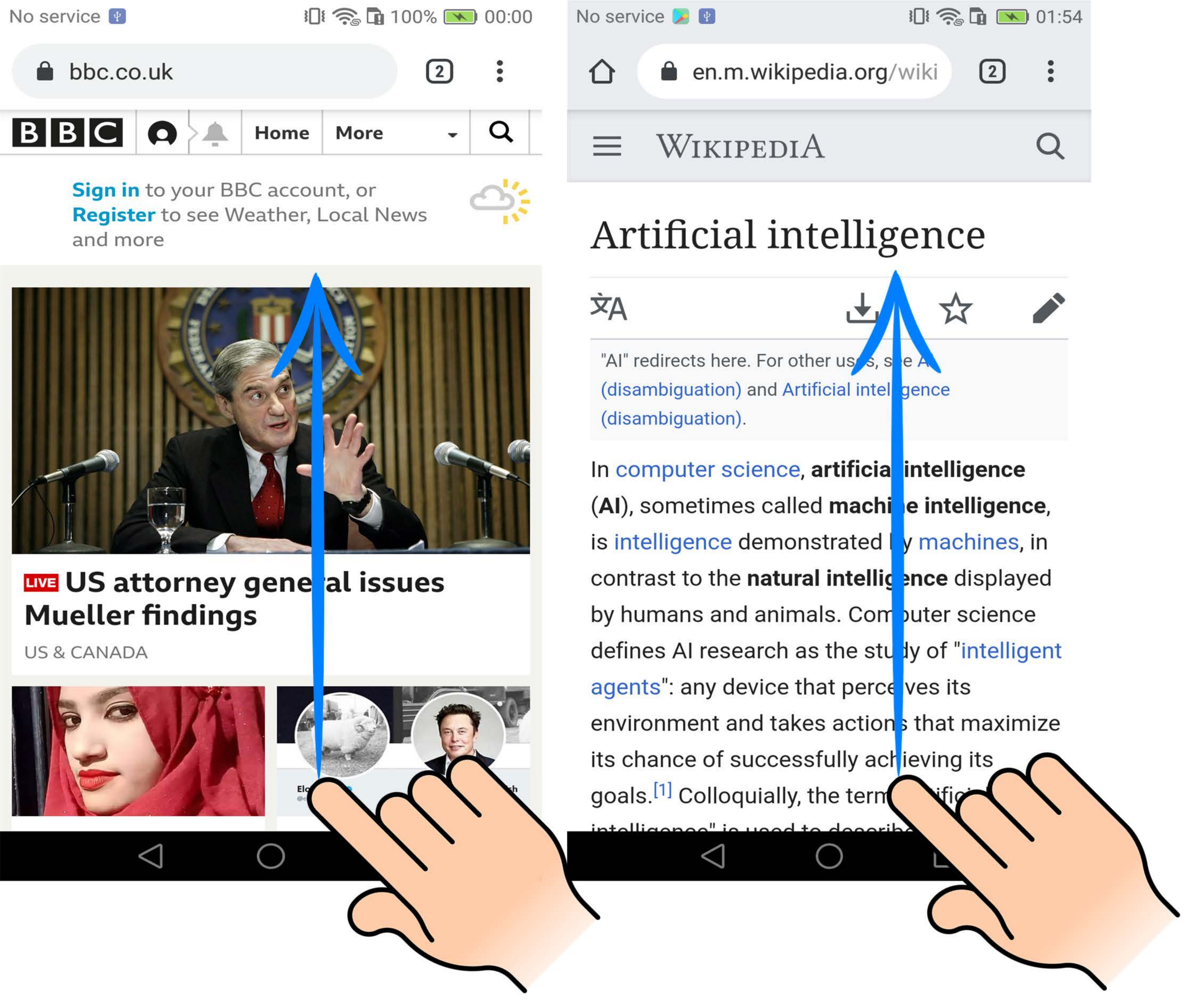}}
    \hfill
    \subfloat[][Energy consumption]{\includegraphics[width=0.22\textwidth]{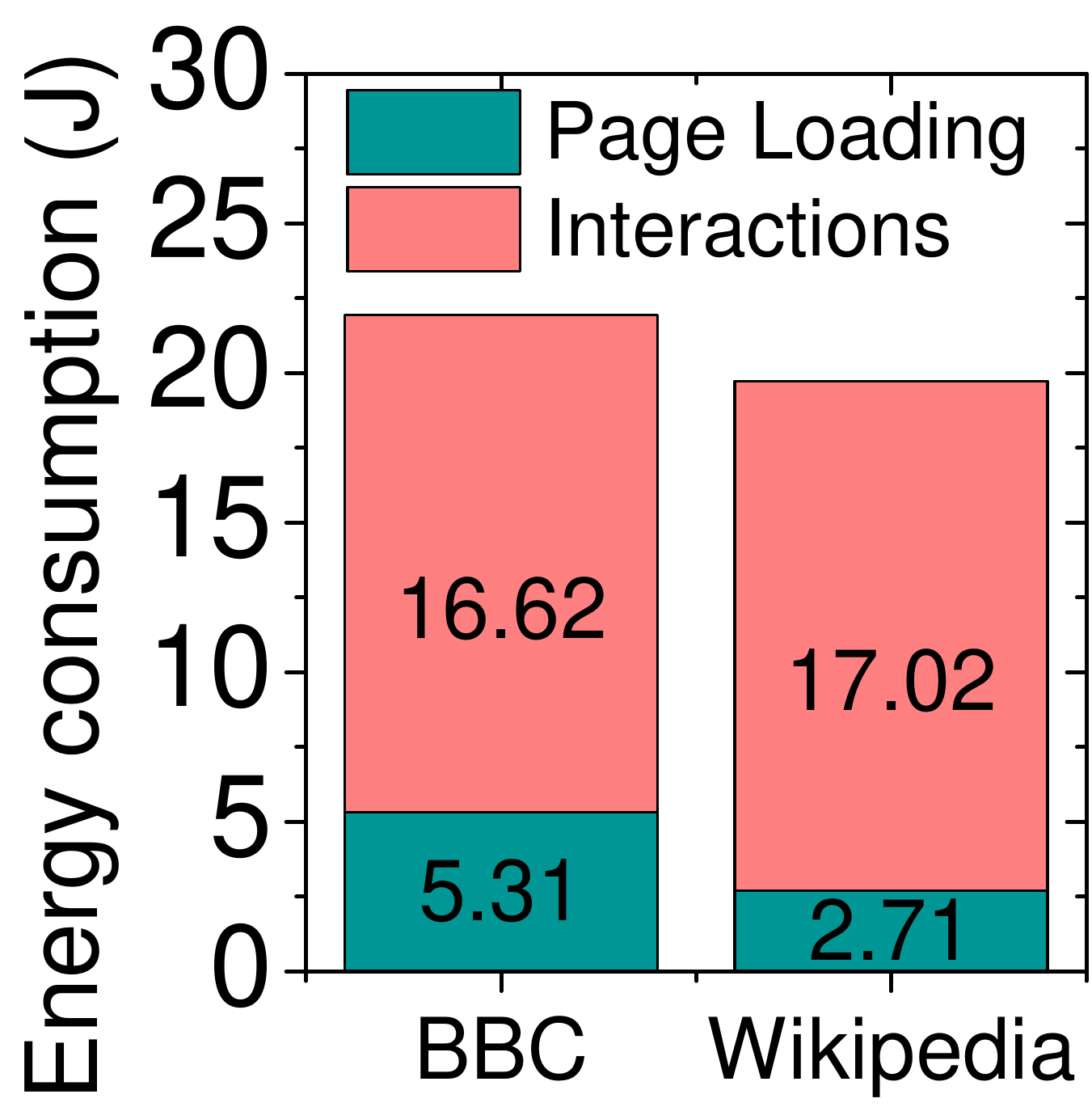}}
    \caption{Motivation webpages (a) and the breakdown of energy consumption during page loading and interactions (b).}
    \label{fig:webview_energy}
\end{figure}

Consider the scenario depicted in Figure~\ref{fig:webview_energy} (a) where a user is scrolling up when reading two webpages from BBC News
and Wikipedia on a recent XiaoMi 9 smartphone (detailed in Section~\ref{sec:setup}). Here, we use RERAN~\cite{gomez2013reran}, a record and
replay tool, to replay user interactions.

\subsubsection{Energy consumption: interactions vs page loading}
Figure~\ref {fig:webview_energy}(b) compares the energy consumed in response to scrolling against that spent during the
loading phase in a WiFi environment. The measurement excludes energy consumption during CPU and GPU idle time. To
minimize the impact of screen use and background workloads, we set the screen to the lowest brightness and close all
background workloads. As can be seen from the diagram, the energy spent during the interaction phase is 2-5 times
higher than that used in the initial loading phase. This finding is in line with prior
studies~\cite{roudaki2015classification,ml2019},  suggesting that prior approaches that only focus on the loading phase
would miss a massive optimization opportunity.

\begin{figure}[!t]
	\centering
	\subfloat[][Energy reduction]{\includegraphics[width=0.23\textwidth]{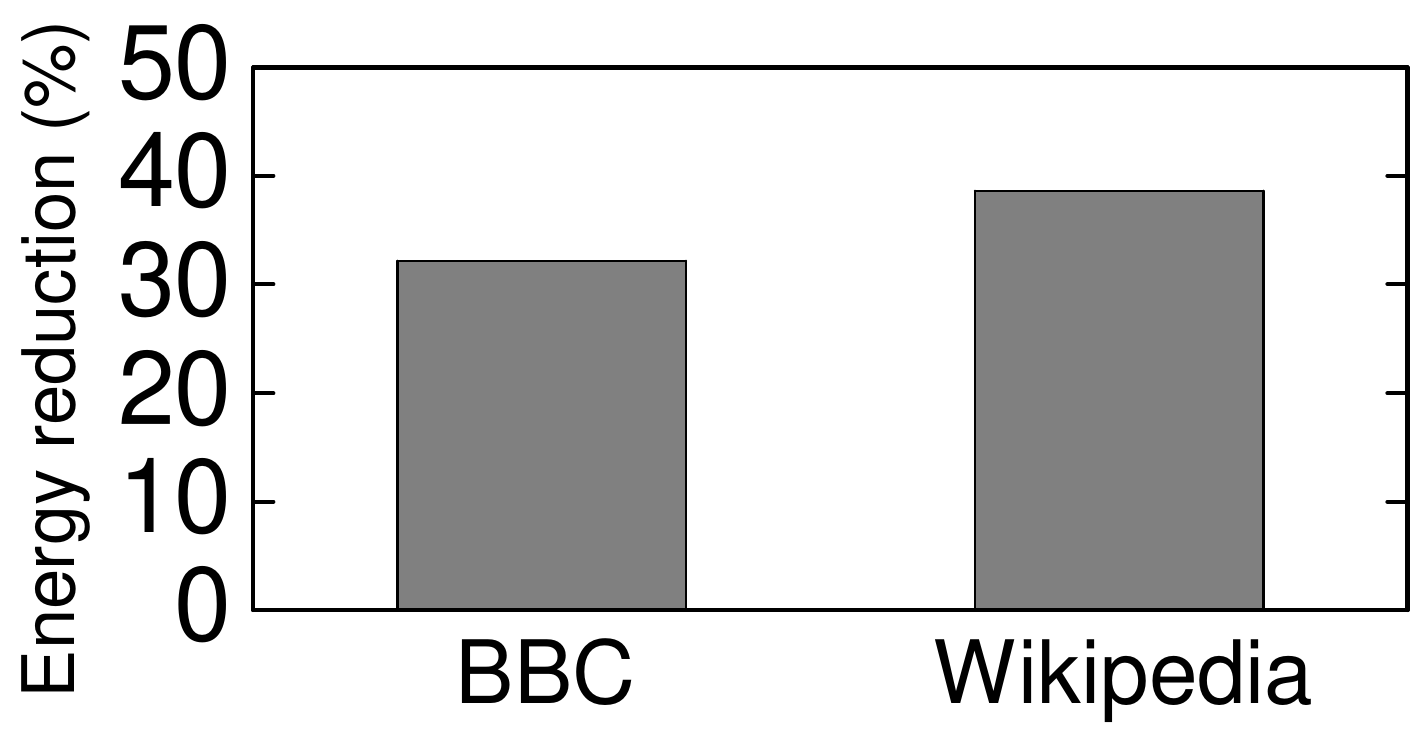}}
    \hfill
    \subfloat[][Frames per second (FPS)]{\includegraphics[width=0.23\textwidth]{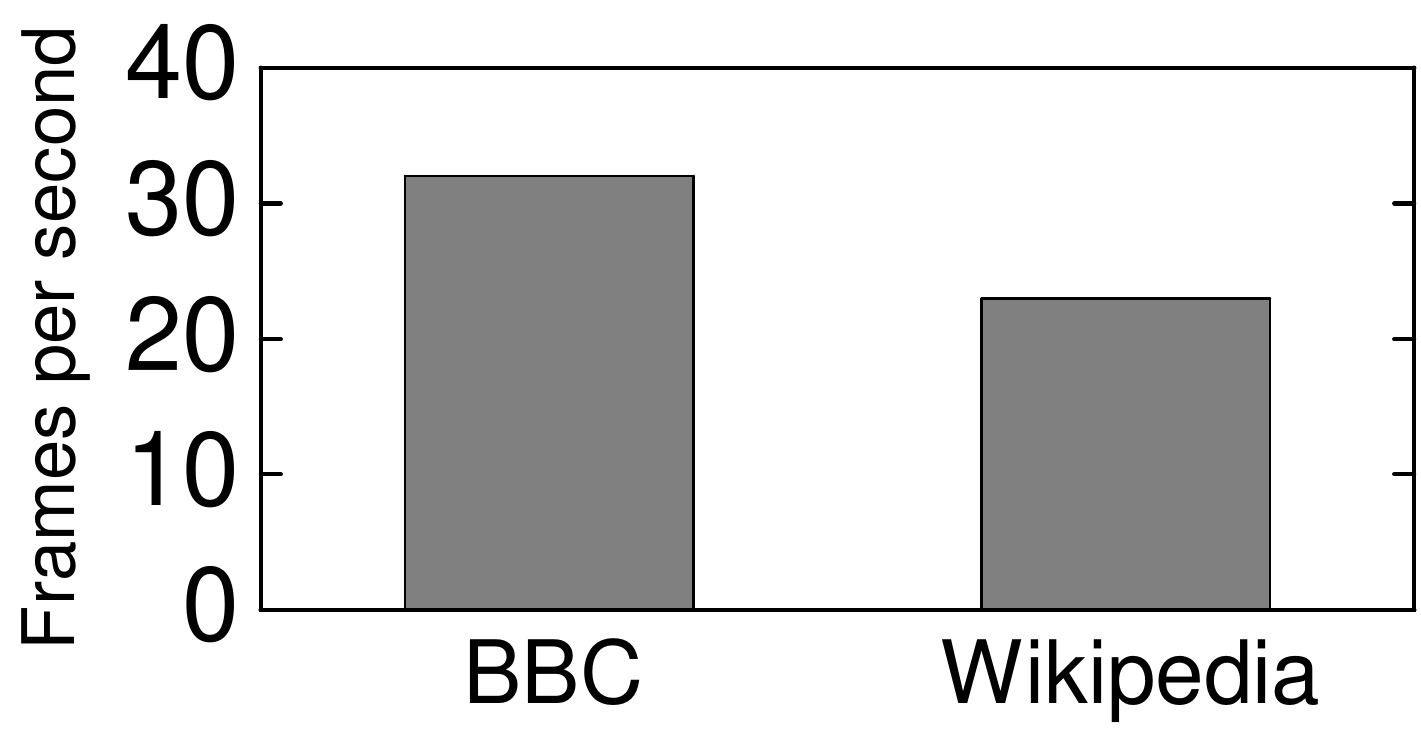}}
    \caption{Energy reduction (a) and FPS (b) on BBC and Wikipedia pages when using the optimal policy over the \texttt{interactive} governor.}
    \label{fig:moti_ux}
\end{figure}

\begin{table}[!t]
\caption{Optimal configurations for motivation examples.}
\scriptsize
\begin{tabularx}{0.5\textwidth}{lXXXX}
        \toprule
		&\textbf{Event-res. freq.}& \textbf{ big CPU (GHz)} & \textbf{little CPU (GHz)} & \textbf{GPU (MHz)}   \\
        \midrule
           \rowcolor{Gray} BBC News    &1 / 6&1.28 &0.672&250\\
             Wikipedia                   &1 / 15&1.05 &0.49&250\\
           \rowcolor{Gray} render process placement      &         &        &\Checkmark& \\
        \bottomrule
\end{tabularx}
\label{tab:best-config}
\end{table}
\subsubsection{Room for improvement}
In the second experiment, we wish to understand how much room is available for trading performance for reduced energy
consumption. We consider two established techniques: (1) setting the CPU/GPU frequency and running the render process
on the big or little CPU cluster, and (2) dropping some of the interaction events (i.e., approximate computing). To
quantify the user expectation, we use frames per second (FPS), because it is shown to strongly correlates to the user's
perceived responsiveness for web browsing~\cite{zhu2015event,ml2019}. For most of the participants in our user study
(see Section~\ref{sec:trainingdata}), the minimum acceptable FPS for the BBC and Wikipedia pages is 32 and 23
respectively. The disparity in the tolerable FPS is due to the content of the two pages. The BBC page is dominated by
images while the Wikipedia one is dominated by dense texts, and human eyes are less sensitive to the change of
text-dominated content~\cite{barten1999contrast}.

Figure~\ref{fig:moti_ux} gives the energy reduction achieved by the optimal  policy over the Android default
\texttt{interactive} frequency governor and the resultant FPS. To find the optimal policy, we automatically replay the
scrolling events and exhaustively profile all possible options. Table~\ref{tab:best-config} lists the best processing
configurations for the testing webpages. The best policy for the BBC page is to respond to one of every six input
events\footnote{Depending on the speed and duration, a gesture often generates multiple events. For example, a flinging
action can trigger over 70 scrolling events.} and run the render process on the little CPU with an appropriate clock
speed for CPUs and the GPU. This configuration reduces energy consumption by 32.2\%. For the Wikipedia page, the best
policy gives an energy saving of 38.6\%. However, applying the best policy of the Wikipedia webpage to the BBC one will
give an FPS of 26 (6 FPS below the target of 32) and a modest energy saving of 2.6\% over the actual optimal policy.
Therefore, simply using one optimal policy found for one webpage to another is likely to either miss optimization
opportunities or compromise QoE.

\subsubsection{Insights}
The two examples show the enormous potential of energy optimization for mobile web interactions. However, finding the right processing
setting is difficult as it depends on the web content, individual user expectation, and hardware. In the next section, we will describe how
\SystemName addresses this challenge by directly modeling the impact of the web content and interactive speed on user acceptable delay
through predictive modeling.

\vspace{-1mm}
\section{Overview of \SystemName}
\vspace{-1mm}

\begin{figure}
\begin{center}
\includegraphics[width=0.45\textwidth]{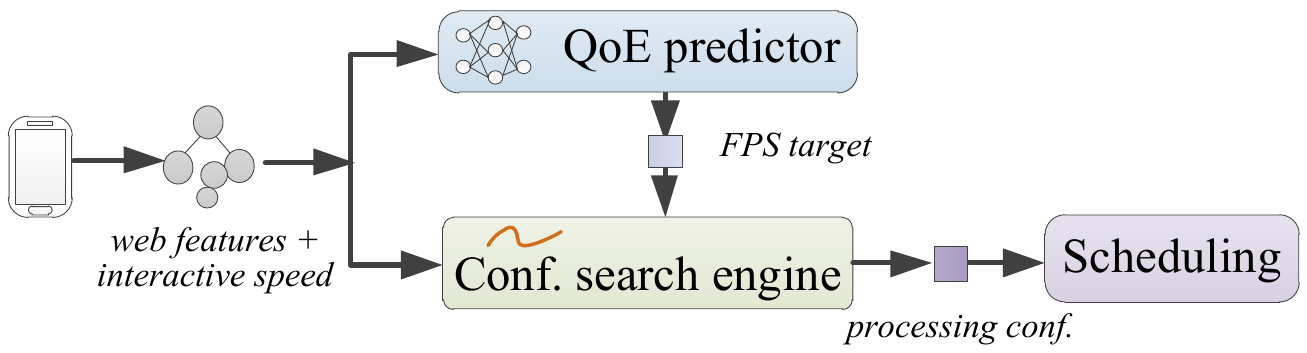}
\end{center}
\vspace{-3mm}
\caption{Overview of the scheduling framework of \SystemName.}
\vspace{-6mm}
\label{fig:overview}
\end{figure}



Figure~\ref{fig:overview}  depicts the scheduling framework of \SystemName. It consists of two innovative components: (a) a QoE predictor
to estimate the minimum acceptable FPS target for a given user, interactive speed and web content, and (b) a configuration search engine to
find a \emph{processing configuration} (i.e., \emph{an event-response frequency and a processor setting}) that meets the minimum FPS
constraint with minimal energy usage.

\vspace{-1mm}
\subsection{QoE Predictor \label{sec:qoepredictor}}
\vspace{-1mm}
Our QoE predictor takes as input features of the web page and the incoming interactive speed. It then predicts the minimum acceptable FPS.
A baseline predictor for each targeting event was first trained ``\emph{at the factory}" through a user study. The baseline predictor then
continuously improves itself for each target user after deployment.

\vspace{-1mm}
\subsection{Configuration Search Engine \label{sec:searchengine}}
Given a content-specific QoE constraint expressed as an FPS target, the configuration search engine finds a processing
configuration to use. This is achieved by using an FPS predictor (or profit estimator) to estimate the FPS as a
function of a processing configuration and web features. By varying the processing configuration given to the
predictor, the search engine can then exam the expected FPS and choose the best-performing configuration before
applying it. The chosen processing configuration is passed to the render-related processes and a runtime scheduler to
adjust the event-response frequency and processor settings. Like the QoE predictor, we learn one FPS predictor for each
event type, three in total.

\vspace{-1mm}
\subsection{Adaptive Learning}
\vspace{-1mm}
\SystemName is designed to be a practical scheme that is portable across users and mobile devices. There are two critical challenges
related to this design goal. Firstly, how to reduce the end-user involvement in capturing a user's QoE requirement. Secondly, how to detect
and improve an ageing decision model in the deployment environment.

To reduce end-user involvement, \SystemName employs transfer learning (Section~\ref{sec:hardwareport}) to quickly re-target an existing
model for a \emph{new} user or platform. Rather than retraining on the entire training dataset,  transfer learning uses only a dozen of
webpages. This not only significantly reduces the profiling overhead but also allows performing learning on the user's device to mitigate
the privacy concern for doing that on a remote server~\cite{xu2018ebrowser}. To detect and improve ageing models, \SystemName uses
conformal predictions (Section~\ref{sec:agining}) to assess the credibility of each prediction. It then uses user feedback or automated
runtime measurements on incorrectly predicted inputs to improve a deployed model over time. This continuous learning strategy minimizes
user intervention by only asking for feedback when things have gone wrong.


\section{Predictive Modeling}
\label{sec:pred}

The QoE and FPS predictors employed by \SystemName are artificial neural networks (ANNs). We choose the ANN because it gives better and
more robust performance over alternatives (Section~\ref{sec:analysis}), and also allows the use of transfer learning to mitigate the
training overhead in the deployment environment (Section~\ref{sec:tf}). We describe our predictive modeling based framework by following
the classical 3-step process for supervised learning: (1) problem modeling and training data generation (2) train a predictor (3) use the
predictor.

\subsection{Problem Modeling and Training Data Generation}
\label{sec:tdata}
\begin{figure}
\begin{center}
\includegraphics[width=0.5\textwidth]{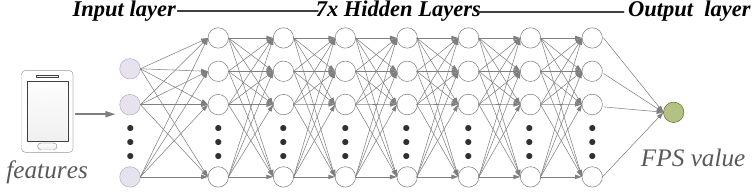}
\end{center}
\caption{Our neural network based predictor.}
\label{fig:ann_structure}
\end{figure}

\subsubsection{Model structure}
Figure~\ref{fig:ann_structure} depicts our neural network - a fully connected, feed-forward ANN with 7 hidden layers and 260 nodes per
hidden layer. The number of nodes of the input layer is determined by the dimensionality of the model input (Section~\ref{sec:features}).
This structure is \emph{automatically determined} by applying the AutoKeras~\cite{jin2018efficient} AutoML tool on the training dataset. In
Section~\ref{sec:analysis}, we evaluate the impact of network structures on performance.

\subsubsection{Training data generation}
\label{sec:trainingdata} We apply cross-validation to train and test our models (see also Section~\ref {sec:methodology}). Training data
are generated by profiling a set of training webpages.

\begin{figure}[t!]
\begin{center}
\includegraphics[width=0.42\textwidth]{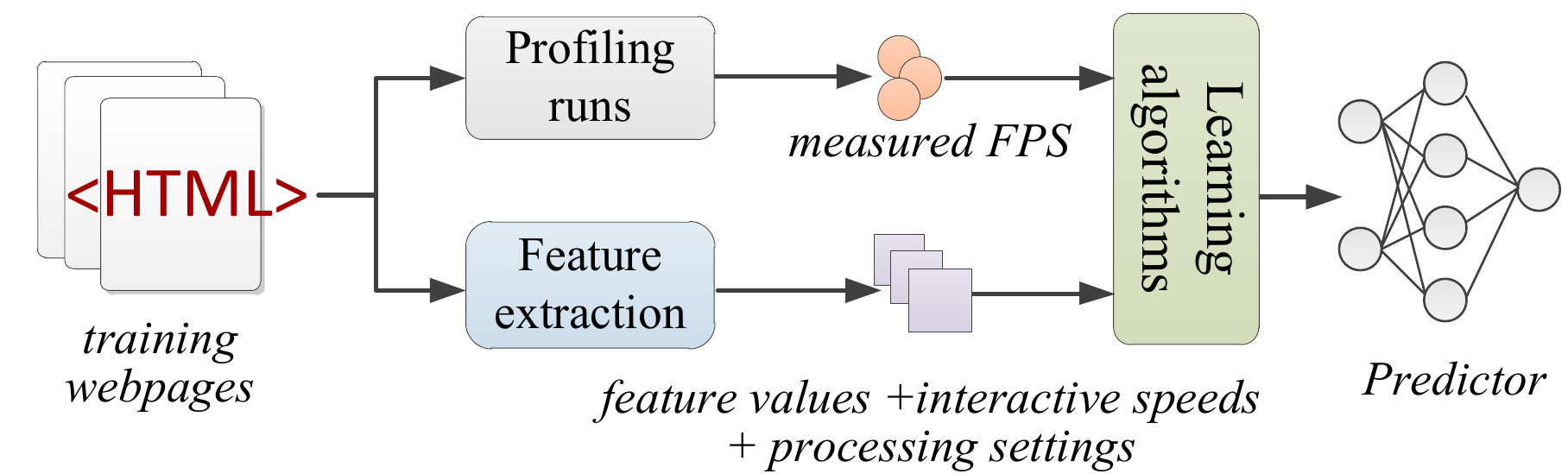}
\end{center}
\caption{Overview of our training process for FPS predictions.}
\label{fig:training_ann}
\end{figure}

\cparagraph{FPS training data.} Figure~\ref{fig:training_ann} depicts the process for learning a baseline FPS predictor on 800 training
webpages. To generate training data, we use RERAN to automatically generate a gesture at different speeds (measured by the number of pixels
per second) on each training webpage. For each interactive speed, we vary the processing configurations and record the achieved FPS.
Specifically, we exhaustively execute the computation-intensive render and paint processes under each CPU/GPU setting. We also evaluate all
candidate event-response frequencies for a processor setting. In total, we train an FPS predictor on over 1 million \emph{automatically
generated} training samples ($800$ webpages $\times$ $10$ interactive speeds $\times$ $\sim16$ processor settings $\times$ $8$
event-response frequencies). The processor settings and event-response frequencies are configurations on the optimal frontier of
performance and energy trade-offs, which are determined by profiling all possible settings on 20 randomly chosen training webpages. Note
that the trained model can be applied to arbitrary interactive speeds and processor settings by taking these as the model inputs. Finally,
for each webpage, we collect the web features as we will described later in this section. We stress that this process is \emph{fully
automated} and does not require user involvement.

\cparagraph{QoE training data.} Our QoE training data are gathered through a user study. In practice, this can be done through a
crowdsourcing platform like Amazon Mechanical Turk. Our user study involved 30 paid users (15 females) who were studying at our
institution. To minimize user involvement, we apply the \emph{k-means} clustering algorithm~\cite{bishop96} to choose 100 representative
webpages from our training dataset.  We ask each user to watch the screen update of each training webpage on a XiaoMi 9 smartphone under
various FPS speeds. We also vary the incoming event by considering 5 commonly interactive speeds per gesture~\cite{xu2018ebrowser} To help
our participants to correlate the generated events to finger movements, we invite them to interact with the device and show the resultant
FPS of their finger movements. For each training instance, we ask a user to select the lowest acceptable screen update rate. We then record
the corresponding minimum acceptable FPS on a per-webpage, per-speed, per-gesture and per-user basis. On average, it took a participant 2.5
hours to complete the study. Later, we extend this user study to all 1,000 webpages used for QoE evaluation using cross-validation.

\begin{table}[t!]
\caption{Raw web features used in the work}
\scriptsize
\centering
        \begin{tabular}{rll}
        \toprule
        \multirow{2}{*}{DOM Tree} & \#DOM nodes & depth of tree \\
                & \#each HTML tag & \#each HTML attr. \\
        \rowcolor[gray]{.92}  & \#rules  & \#each property \\
        \rowcolor[gray]{.92}  \multirow{-2}{*}{Style Rules} & \#each selector pattern & \\
        Other  & GPU memory footprint for viewports & \\
        \bottomrule
        \end{tabular}
\label{tab:feature}
\end{table}

\subsubsection{Feature extraction}
\label{sec:features} One of the key aspects in building a good predictor is finding the right features to characterize the input workload.
In this work, we started from 6,913 raw web features extracted from the DOM tree. Table~\ref{tab:feature} summarizes our raw features. The
features were chosen based on previous work of mobile web optimization~\cite{Ren:2018:PNW:3281411.3281422} and our intuitions.

The QoE model takes as input the web features of current and future viewports and the user interactive speed. The FPS model takes as input the web
features, the interactive speed, the processing setting (i.e., \emph{event-response frequency and processor setting}), and the CPU cluster where
the render process is running on (for modeling the penalty for cross-processor task migration).

\cparagraph{Feature reduction.} To learn effectively over a small training dataset, we apply the correlation coefficient and principal
component analysis~\cite{bishop96} to reduce the dimensionality of raw web features from 6,913 to 127. Both techniques are shown to be
useful in prior work for feature reduction~\cite{Ren:2018:PNW:3281411.3281422,pes}.

\cparagraph{Feature normalization.} In the final step, we scale each of the extracted feature values to a common range
(between 0 and 1) to prevent the range of any single feature being a factor in its importance. We record the minimum
and maximum values of each feature in the training dataset, in order to scale the feature values of an unseen webpage.
We also clip a feature value to make sure it is within the expected range during deployment.

\subsubsection{Training overhead}
The time for training the baseline predictors is dominated by generating the training data. In this work, it takes less than a week to
collect all the training data for a mobile platform. In comparison processing the raw data, and building the models took a negligible
amount of time, less than an hour for learning all individual models on a PC. We stress that training of the baseline predictors is a
one-off cost.

\subsection{Training a Baseline Predictor}

\label{sec:training} The collected web feature values and speed together with the desired FPS values are passed to a supervised learning
algorithm to learn an ANN for each event. For FPS predictions, we also use additional model inputs as stated in Section~\ref{sec:features}.
Our models are trained using back-propagation with stochastic gradient descent (SGD) guided by the widely used Adam optimizer~\cite{kingma2014adam} and L2 regularization, which is a standard-setting for training ANNs.  For training examples $y_1 \ldots y_n$, the
optimizer finds model parameters $\Theta$ to minimize the output of a mean-squared-logarithmic loss (MSLE) function $\ell$:

{\scriptsize
$$
\Theta = \argmin_{\Theta} \frac{1}{n} \sum_{i=1}^{n} \ell \left(y_i, \Theta \right)
$$
}

We choose MSLE because it penalizes underestimates more than overestimates, which reduces the chance of QoE violations due to an
underestimated FPS target.

\subsection{Using the Models}

The trained predictors can be applied to \emph{new, unseen} webpages. We implemented our models using Keras~\cite{keras} and
Scikit-learn~\cite{scikitlearn}. Our optimization can be turned on by the user or enabled during Android's ``Battery Saver" mode. When a
supported event is detected, we will extract web features of the current and the future viewports from the DOM tree -- the future viewport
is calculated based on the interactive speed. We calculate the average interactive speed using a sampling window of 50 ms or the
interactive session -- which is shorter. We then use the QoE and FPS predictors to choose the optimal processing configuration as described
in Section~\ref{sec:searchengine}. To minimize the runtime overhead, our framework runs on the least loaded CPU core. The overhead of
feature extraction, predictions, searching, and runtime scheduling is small -- less than 5 ms, which is included in all experimental
results.


\section{Adaptive Learning}
\label{sec:adapt} 

 We propose two new ways to improve the adaptiveness and practicability of a machine-learning-based web optimizer.

\subsection{Adapt to A New Environment}
\label{sec:hardwareport}

\subsubsection{The problem}
QoE is user-specific and the resultant FPS depends on the underlying hardware. Therefore, using a generic model across
different users and hardware platforms is ineffective. To tune a model to match a specific user or mobile device,
\SystemName employs transfer learning~\cite{long2017deep} to quickly port a baseline predictor to the target computing
environment.


\subsubsection{The idea} Prior work in other domains has shown that ANN models trained on similar inputs for
different tasks often share useful commonalities\cite{hu2018learning}. Our work leverages this insight to speed up the process for tuning a
model for a new user or mobile hardware. This is because the first few layers (i.e., those close to the input layer) of our ANN are likely
to focus on abstracting web features and largely independent of the model output. Since we use the same network structure, transfer
learning is achieved by copying the weights of a baseline model to initialize the new network. Then, we train the model as usual but using
profiling information (as described in Section~\ref {sec:tdata}) collected from fewer training webpages.

\subsubsection{Determining training samples\label{sec:training_samples}}

A key question for applying transfer learning in our context is how many training examples do we need. Under-provisioning of training data
will lead to low accuracy, while over-provisioning will incur significant profiling overhead especially when that requires end-user
involvement. To determine the right number of training examples, we group our training webpages using the \emph{k-means} clustering
algorithm. We then choose two webpages from each cluster: one is mostly close to its cluster centroid on the feature space, the other has
the biggest Frobenius norm value~\cite{bauckhage2015k} with respect to other centroid points. In practice, the chosen webpages can be
shipped as part of the browser bundle, where profiling can be performed when the device is charging after the first installation.

To determine the right number of clusters (i.e., $K$), we use the \emph{Bayesian Information Criterion} (\texttt{BIC}) score~\cite{657808}.
The BIC measures if a selected $K$ is the best fit for grouping data samples within a dataset. The larger the score is, the higher the
chance that we find a good clustering number for the dataset. The BIC score is calculated as~\cite{605403}:

{\scriptsize

$$
BIC_j=\hat{l}_j-\frac{p_j}{2}\cdot logR
$$

} where $\hat{l}_j$ is the likelihood of the data when $K$ equals to $j$, $R$ is the number of training samples, and the free parameter
$p_j$ is the sum of $K-1$ class probabilities -- calculated as: $p_j=(K-1) + dK + 1$ for a $d$-dimension feature vector plus 1 variance
estimate. $\hat{l}_j$ is computed as: {\scriptsize
$$
\hat{l}_j=\sum_{n=1}^{k}-\frac{R_n}{2}log(2\pi)-\frac{R_n\cdot d}{2}log(\hat{\sigma}^2) -\frac{R_n-K}{2} +R_nlog(R_n/R)
$$
}
where $R_n$ is the number of points in a cluster and $\hat{\sigma}^2$ is the distance variance between each point to its cluster
centroid.

\begin{figure}
\begin{center}
\includegraphics[width=0.35\textwidth]{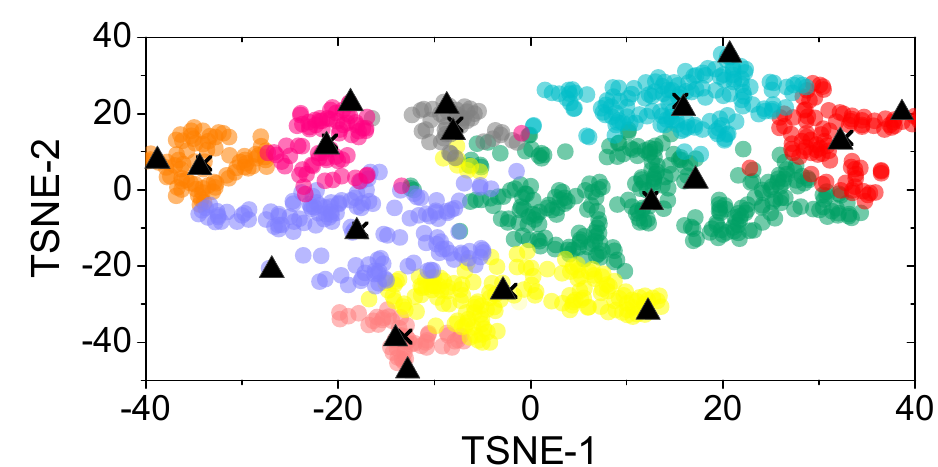}
\end{center}
\caption{Using clustering to choose training examples for transfer learning. A cluster centroid is
marked by a cross, while the two chosen webpages of a cluster are marked as triangles.} \label{fig:tsne}
\end{figure}

\subsubsection{Illustrative example}
Figure~\ref{fig:tsne} illustrates how one of our training dataset of 800 webpages can be grouped into 9 clusters
determined using the BIC score. To aid the clarity, we apply t-SNE~\cite{maaten2008visualizing} to project the data
onto a two-dimensional space. Directly using an FPS model trained for Pixel 2 to XiaoMi 9 gives an error rate of
37.5\%. By using profiling information collected from 18 carefully chosen webpages on the target device to update the
predictor, the error rate decreases to 6.7\%. Such performance is not far from the error rate of 4.6\% when training
the model from scratch by profiling the entire 800 training webpages, but we cut the training time from two days to
less than two minutes on the end user's phone.

\subsection{Continuous Learning at the Deployment Environment}
\label{sec:agining}

\subsubsection{The problem} 
The key for continuously improving a model after deployment is knowing when the model is wrong so that we can use the ground-truth to
improve it. Judging if an FPS prediction is inaccurate is straightforward because the ground-truth can be automatically measured. Checking
if a QoE target meets the user expectation is harder because we cannot ask a user to provide feedback every time.

\subsubsection{The solution} To estimate if a QoE target prediction is wrong, we leverage the conformal prediction (\texttt{CP})~\cite{shafer2008tutorial,balasubramanian2014conformal}. The CP
is a statistical assessment method for quantifying how much we could trust a model's prediction. This is done by learning a
\emph{nonconformity function} from the model's training data. This function estimates the ``strangeness" of a mapping from input features,
$x$, to a prediction output, $y$, by looking at the input and the probability distribution of the model prediction. In our case, the
function estimates the error bound of a QoE prediction. If the error bound is greater than a configurable threshold (20\% in this work), we
then consider the model gives an incorrect prediction.

\SystemName uses the inductive CP as it works with any regression model~\cite{volkhonskiy2017inductive}. For a prediction, $y$, of input
$x$, function $f$ calculates the nonconformity score as:

{\scriptsize
$$
f (x, y) = \frac{|y-h(x)|}{g(x) + \beta}
$$
} where $h$ is a regression-based QoE or FPS model, $g$ estimates the difficulty of predicting $y$ and $\beta$ is a sensitive parameter
that determines the impact of normalization. Note that $g$ and $\beta$ are automatically determined from training data.

\subsubsection{Continuous learning}
For a QoE prediction that is considered to be inaccurate, \SystemName takes the high-end value of the CP-estimated error bound to minimize
QoE violations. It then finds out the actual QoE target by seeking user feedback. This is done by automatically replaying the screen update
under different FPS settings, from high to low. For each setting, \SystemName asks the user to rate the screen update for being
``acceptable" or not. It stops playing the screen update when the user indicates an FPS setting is unacceptable. To update a QoE or FPS
model, \SystemName adds profiling information of the uncertain inputs to the transfer learning dataset. When the device is charging,
\SystemName runs the learning algorithm to update the predictors and CP models.


\section{Evaluation Setup\label{sec:setup}}

\begin{table}[t!]
\begin{center}
\caption{Evaluation platforms}
\vspace{-2mm}
\label{tbl:platforms}
\scriptsize
\begin{tabularx}{0.49\textwidth}{p{1cm}p{1.7cm}Xp{0.6cm}p{0.6cm}X}
\toprule
\textbf{Device} &\textbf{CPU} & \textbf{GPU} & \textbf{RAM (GB)} & \textbf{Screen (inches) } & \textbf{OS}\\
\midrule
\rowcolor{Gray} XiaoMi 9 & Snapdragon 855 @ 2.84 GHz & Adreno 640 & 8 & 6.39 &  MIUI 10 (Android 9)\\
Google Pixel 2 & Snapdragon 835 @ 2.35 GHz &  Adreno 540 & 4 & 5.0 & Android 9 \\
\rowcolor{Gray} Huawei P9 & Kirin 955    @ 2.5 GHz & Mali T880  & 3 & 5.2 & Android 8\\
Odroid Xu3 & Exynos 5422   @ 2 GHz &  Mali T628 & 2 & 4 & Ubuntu 16.04\\
\bottomrule
\end{tabularx}
\end{center}
\end{table}

\begin{figure}[!t]
	\centering
	\subfloat[][\#DOM nodes]{\includegraphics[width=0.23\textwidth]{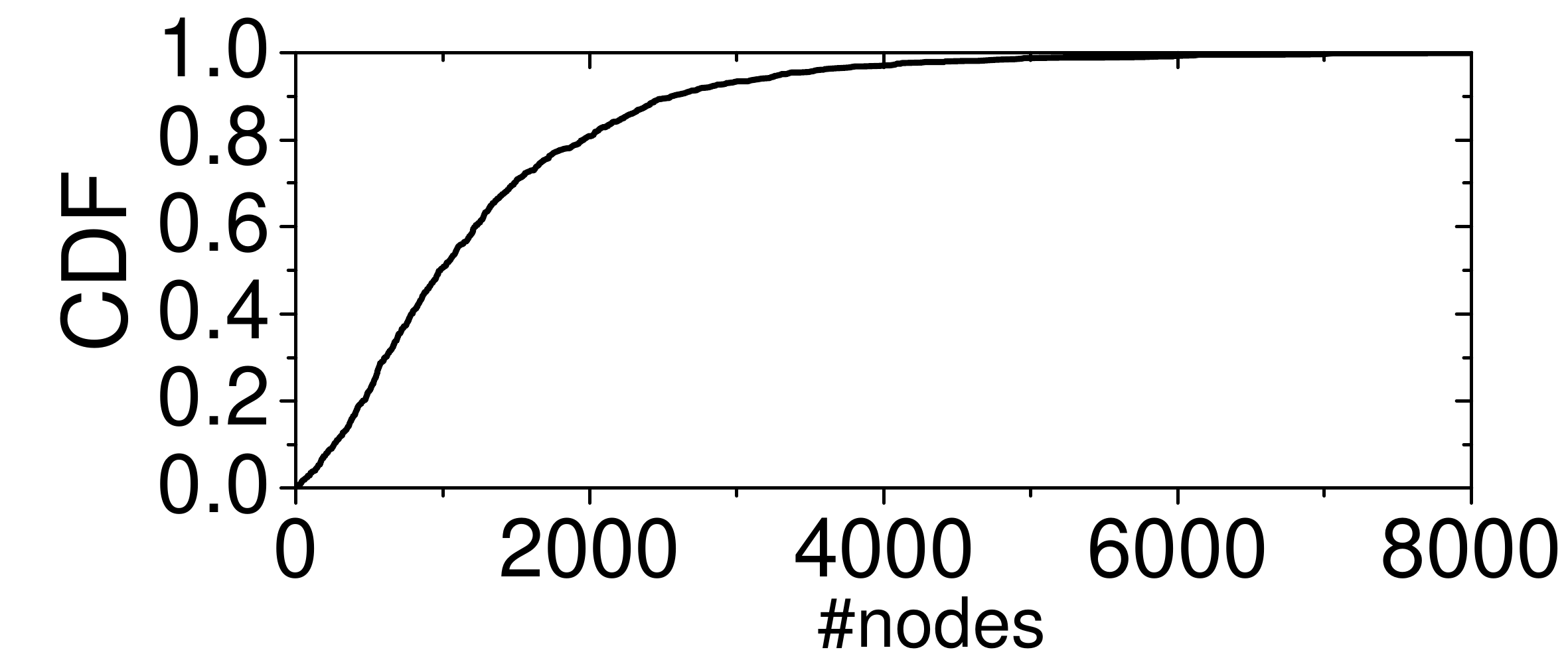}}
    \hfill
    \subfloat[][Webpage size]{\includegraphics[width=0.23\textwidth]{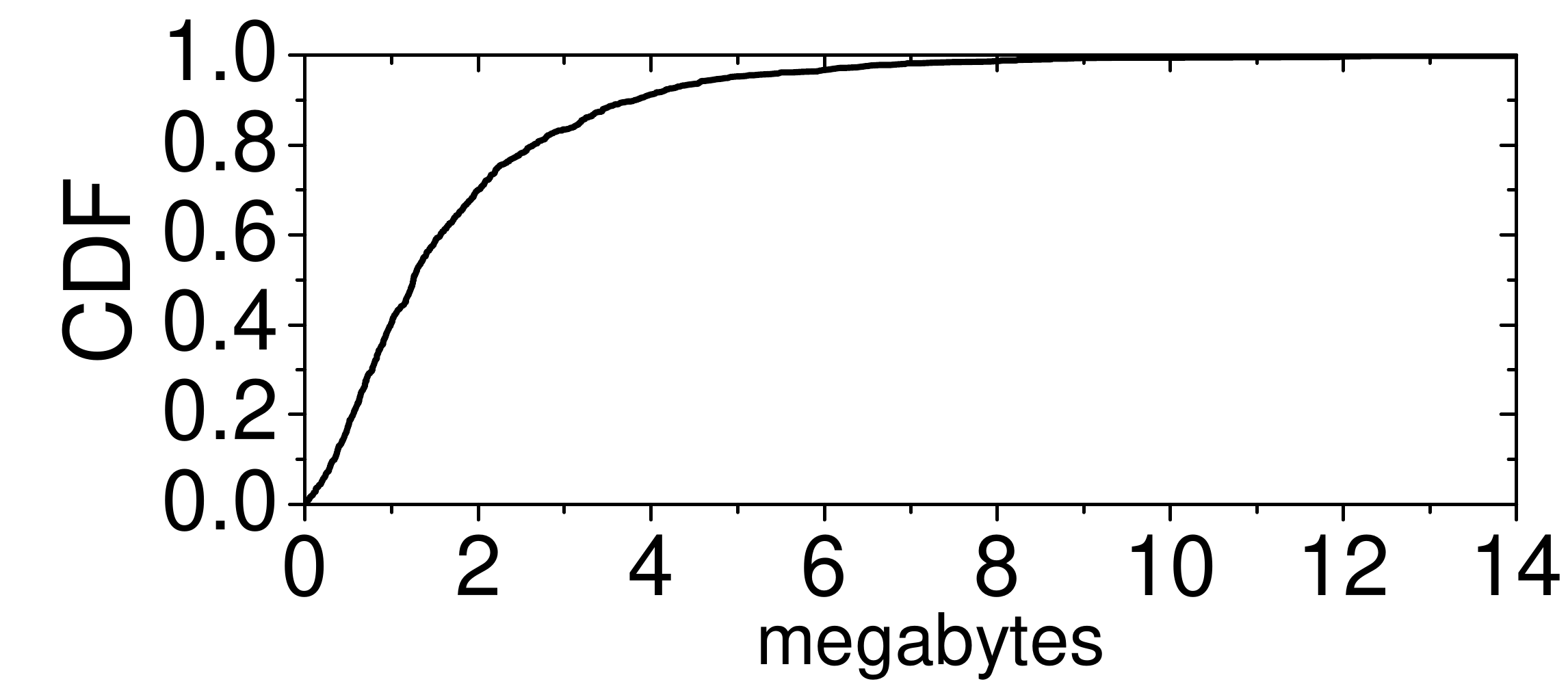}}
    \caption{The CDF of \#DOM nodes (a), webpage size (b).}
    \label{fig:web-workloads}
\end{figure}

\subsection{Platforms and Workloads}

\cparagraph{Evaluation Platforms.} To implement \SystemName, we modified Chromium (ver. 74)\footnote{Code can be
downloaded from \emph{[https://bit.ly/2srZbs9]}.} and compiled it under the ``release" build.  Our evaluation
platforms, detailed in Table~\ref{tbl:platforms}, include different hardware specs, representing low, medium and
high-end mobile systems. We specifically include Odroid Xu3, because all, except one~\cite{xu2018ebrowser}, of our
competitive schemes have been tuned and evaluated on this platform.

\cparagraph{Web Workloads.} We use the landing page of the top 1,000 hottest websites (as of May, 2019) ranked by \texttt{alexa.com} based
on the global  web traffic analysis.  Figure~\ref{fig:web-workloads} shows the CDF of the number of DOM nodes and web content sizes. The
webpage sizes range from small (4 DOM nodes and 10 KB) to large (over 7000 DOM nodes and 14 MB), indicating that our test data cover a
diverse set of web contents.

\vspace{-1mm}
\subsection{Competitive Approaches}
\vspace{-1mm}
\label{sec:comps} We compare \SystemName to the following state-of-the-arts:

\begin{itemize}

\item \texttt{EBS}: A regression-based method for adjusting the processor frequency to meet a fixed response
    deadline~\cite{zhu2015event};

\item \texttt{Phase-aware}: An event-phase-based power management strategy for mobile web browsing~\cite{phaseaware};

\item \texttt{ML-governor}: A machine-learning-based CPU frequency governor for interactive web browsing~\cite{ml2019};

\item \texttt{eBrowser}: This strategy puts the browser process into sleep to drop some of the input user events~\cite{xu2018ebrowser}.

\end{itemize}

All the above schemes require learning on the entire training dataset for each hardware architecture. Moreover, all, except eBrowser,
assume a fixed deadline for an event type.

\subsection{Evaluation Methodology}
\label{sec:methodology} \label{sec:method}

\cparagraph{Model evaluation.} Like~\cite{ml2019}, we use five-fold cross-validation to train all machine learning models (including our
competitors). Specifically, we randomly partition our 1,000 websites into 5 sets where each set contains webpages from 200 sites. We keep
one set as the validation data for testing our model, and the remaining 4 sets as training data to learn a model. We repeat this process
five times (folds) to make sure that each of the 5 sets used exactly once as the validation data. To minimize user involvement, we use a
subset of webpages from the training dataset to build the QoE model as described in Section~\ref{sec:trainingdata}.  This is a standard
methodology for evaluating the generalization ability of a learned model.

\cparagraph{Metrics.} We consider two metrics: energy saving and QoE violation. Energy saving is normalized to the energy consumed by the
\texttt{interactive} scheduler, an Android default CPU governor for interactive applications. QoE violation is calculated as
$\delta/FPS_{min}$, where $\delta$ is the number of FPS falls below the minimum acceptable FPS, $FPS_{min}$~\cite{ml2019}. We do not use
\texttt{powersave} as a baseline as it gives long processing times and violates QoE for all our test cases.

\cparagraph{Measurements.} For energy measuring, we use a Monsoon power meter~\cite{moonsoon} (except for Odroid Xu3 because it already has
onboard power sensors for energy measurement) to measure the power consumption of the \emph{entire} system including the display with a
50\% brightness (a typical indoor setting of Android). For the FPS, we use a script to count the number of invocations of the
\texttt{SurfaceView} object of Chromium.

\cparagraph{Reporting.} When reporting performance, we use the \emph{geometric mean}, which is widely seen as a more reliable performance
metric over the arithmetic mean~\cite{ertel1994definition}. Unless state otherwise, we report the geometric mean across 3.6 million
automatically-generated test cases of 1,000 webpages, 30 users, 4 devices, 3 gestures and 10 speeds per gestures, using
cross-validation. Moreover, events are automatically generated, starting from the initial viewport of a webpage.  To have statistically
sound data, we run each approach on a test case repeatedly until the confidence-bound under a 95\% confidence interval is smaller than 2\%.
Finally, all webpages are loaded
from the device's internal storage to preclude network variances, and we disable the browser cache to ensure consistent results across runs. 

\section{Experimental Results}

\begin{figure}[t]
	\centering
	\subfloat[][Energy reduction]{\includegraphics[width=0.23\textwidth]{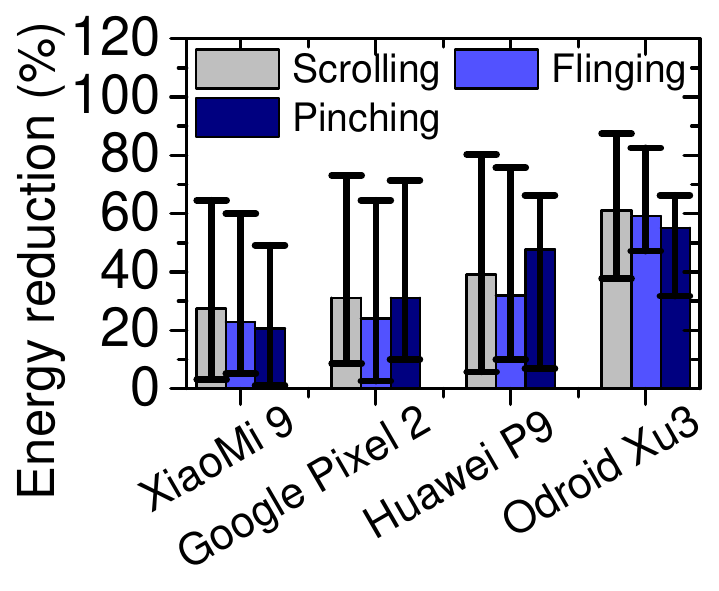}}
    \subfloat[][QoE violation]{\includegraphics[width=0.22\textwidth]{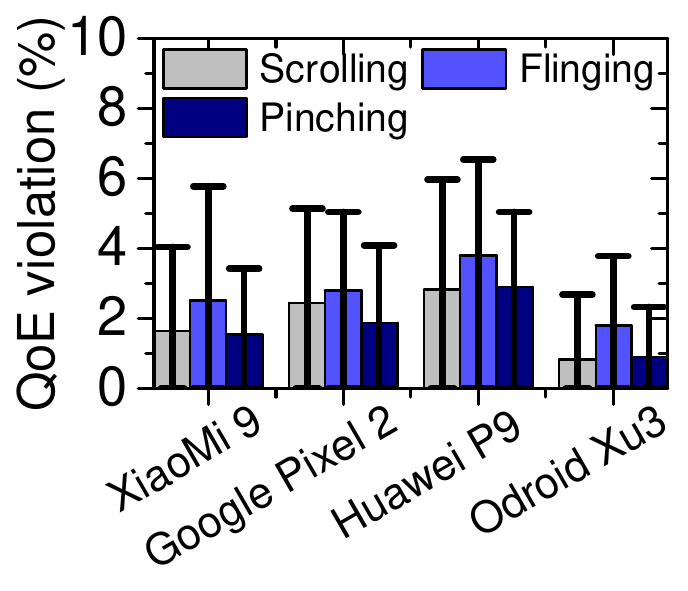}}
    \caption{The energy reduction (a) and QoE violations (b) achieved by our approach over \Interactive.}
    \label{fig:overallresult}
\end{figure}

\subsection{Content-aware QoE Optimization}
\label{sec:qoep}  To evaluate the benefit of content-aware QoE optimizations, in this experiment we train our predictors on the entire
training dataset, but we will evaluate transfer learning in Section~\ref{sec:tf}. The results are given in Figure~\ref{fig:overallresult},
where the min-max bars show the variances across our evaluation scenarios.

Figure~\ref{fig:overallresult}a shows that \SystemName reduces energy consumption by at least 23.6\% (up to 58.5\%), and
Figure~\ref{fig:overallresult}b confirms that such large energy reduction does not necessarily come at the cost of poor user experience.
\SystemName only leads to 1\% to 4\% of QoE violations on less than 5\% of the testing webpages with 2 to 3 lower than expected FPS values.
On testing webpages where no QoE violation occurred, \SystemName delivers 92.4\% of the available energy savings given by a
\emph{theoretically perfect} predictor (found by exhaustively profiling) that always chooses the optimal processing configuration.
Furthermore, if we take a conservative approach by adding 10\% to the predicted FPS QoE target, \SystemName can then eliminate all the QoE
violations, but still gives an average energy reduction of 21.3\% (12.1\% to 37.4\%). This results show that \SystemName is highly
effective in trading performance for energy savings.

\subsection{Compare to Competitive Approaches}
\begin{figure}[t]
	\centering
	\subfloat[][Energy reduction]{\includegraphics[width=0.23\textwidth]{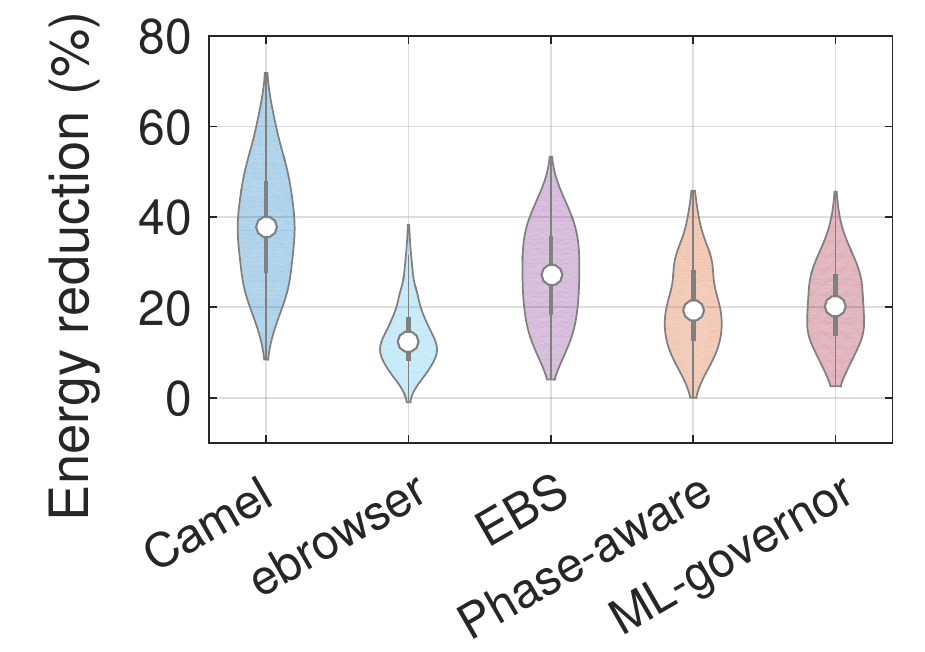}}
    \subfloat[][QoE violation]{\includegraphics[width=0.23\textwidth]{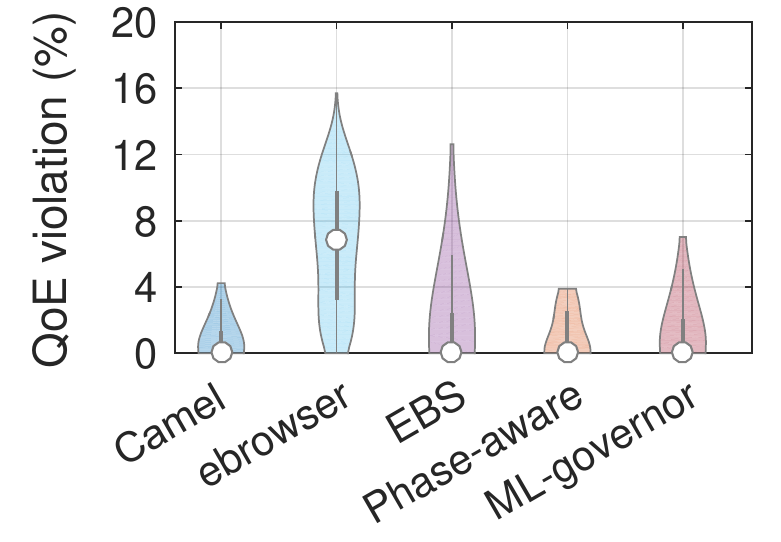}}
    \caption{Compare with the state-of-the-arts.  \SystemName consistently outperforms all alternatives.}
    \label{fig:competitive}
\end{figure}

Figure~\ref{fig:competitive} compares \SystemName with alternative schemes. The white dot in the plots denotes the median value and the
thick black line represents 50\% of the data. For fair comparison, all schemes are built from the same training dataset.

All approaches improve over the \Interactive baseline. By modeling the impact of web content on QoE and using this to configure the
heterogenous hardware, \SystemName gives the highest overall energy saving and the lowest QoE violation ratio. Specifically, \SystemName
reduces the energy consumption by at least 14.6\% (up to 29\%), but with at least 25.1\% (up to 88.3\%) lower QoE violations compared to
prior methods.

\begin{figure*}[t]
	\centering
	\subfloat[][Error rate]{\includegraphics[width=0.24\textwidth]{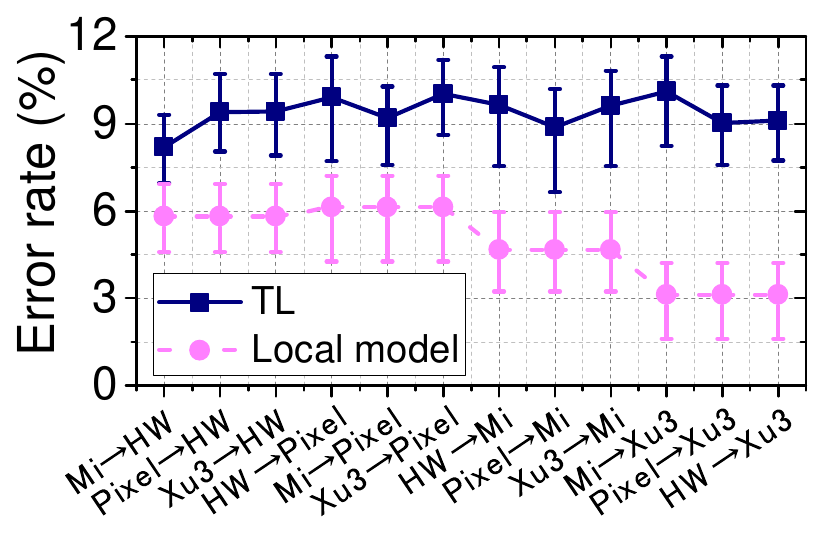}}
    	\subfloat[][Energy reduction]{\includegraphics[width=0.25\textwidth]{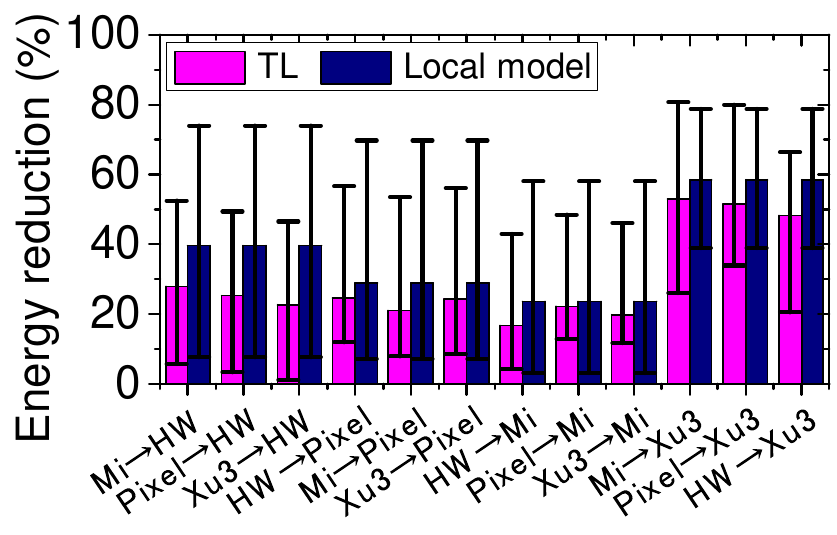}}
    	\subfloat[][QoE violation]{\includegraphics[width=0.24\textwidth]{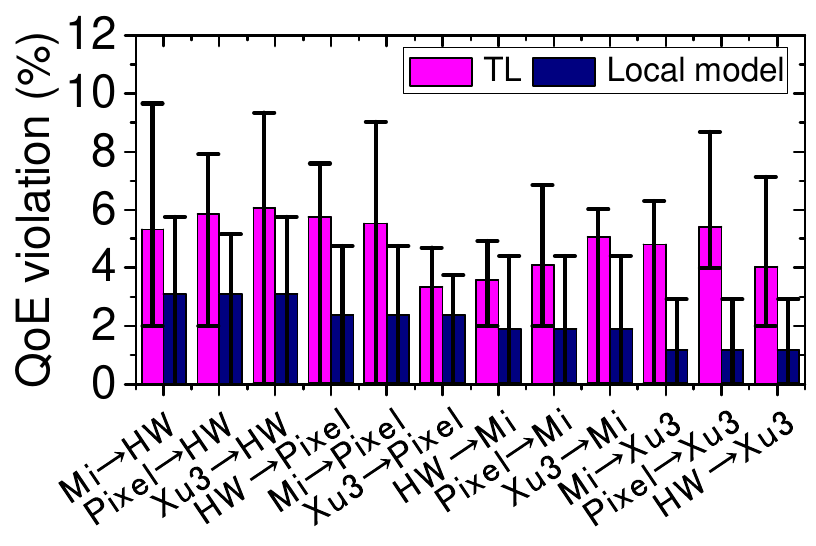}}
    \subfloat[][From Mi9 to HW P9]{\includegraphics[width=0.24\textwidth]{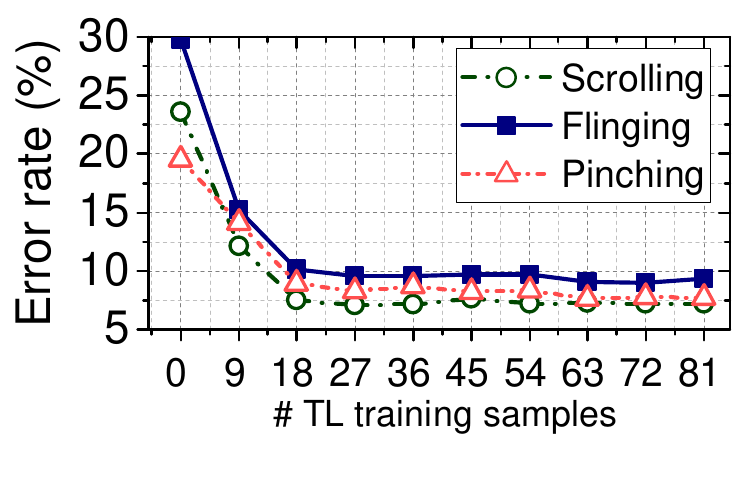}}
    \caption{Transfer learning across platforms. We can use profiling information collected from 18 webpages to update a model. }
    \label{fig:tl_platform}
\end{figure*}

\begin{figure}[t]
	\centering
	\subfloat[][TL within and across groups]{\includegraphics[width=0.24\textwidth]{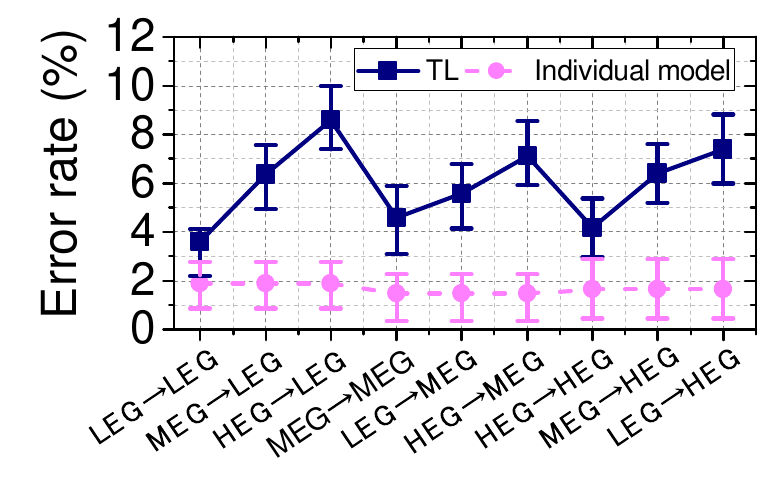}}
    \subfloat[][Perf. w.r.t. \#webpages]{\includegraphics[width=0.24\textwidth]{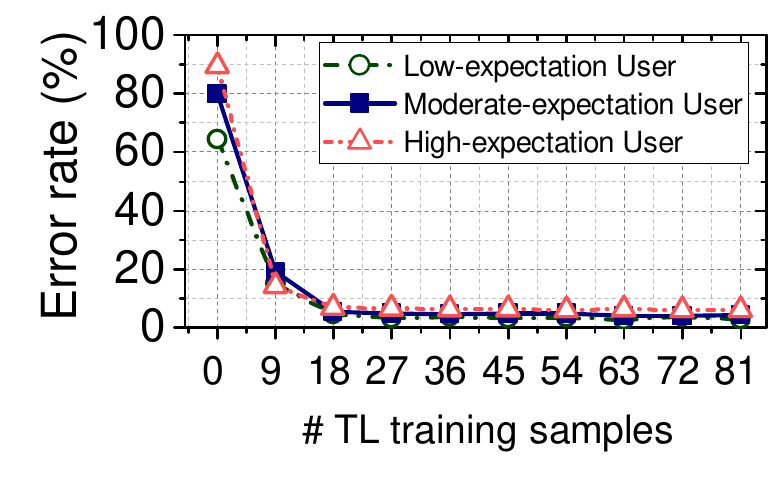}}
    \caption{Applying transfer learning for QoE predictions.}
    \label{fig:tl_qoe}
\end{figure}

\subsection{Evaluation of transfer learning}
\label{sec:tf} We now evaluate our strategy for applying transfer learning (TL) to tune baseline predictors for a new environment. On
average,  TL delivers 27.4\% (up to 53.1\%) of energy savings with less than 6\% of QoE violations. This performance is comparable to the one
reported in Section~\ref{sec:qoep} when the QoE and FPS predictors are trained from scratch every time.

\subsubsection{Tuning FPS predictors} Figure~\ref{fig:tl_platform} shows the results for using TL to port a baseline FPS predictor for a
new platform. Although we only use 2.3\% of the training examples (i.e., 18 webpages - see
Section~\ref{sec:training_samples}), performance of the TL-learnt model is compared to training a completely new model
using 800 webpages. We see only a marginal increase of 4.46\% in the error rate. As subgraphs b and c in
Figure~\ref{fig:tl_platform} show, on average, TL gives 29.7\% of energy reduction with 4.9\% of QoE violations for
porting an FPS predictor to a new platform.

Figure~\ref{fig:tl_platform}d shows how the error rate changes as we increase the number of training webpages when using TL to port an FPS
model built for XiaoMi 9 to Huawei P9. Using more webpages does improve prediction accuracy. However, the performance reaches a plateau
when using 18 webpages, and a further increase in the number of training webpages does not justify the increased profiling overhead. 

\subsubsection{Tuning QoE predictors} We divide the 30 participants of our user study into 3 groups based on their minimum acceptable FPS. The
low-expectation group has 10 users with an averaged FPS target of under 35; the moderate-expectation group  has 14
users with an averaged FPS target of between 35 and 49; and the high-expectation group has 6 users with an averaged FPS
target of over 49.

Figure~\ref{fig:tl_qoe}a reports the performance for applying TL (with cross-validation) to port a QoE predictor to another user from the
same or a different group. As expected, TL within the same user group gives the lowest error rate of between 3.1\% (1.1 FPS) and 4.58\%
(2.08 FPS). We see a slight increase in the error rate when applying TL across user groups, but the average error rate is 6.9\% (2.94 FPS).
In practice, we can further improve the performance by choosing a pre-trained model that is as close as possible to the target user based
on observations seen from the first few webpages, e.g., using a recommendation
system~\cite{bell2008bellkor,Hsieh:2017:CML:3038912.3052639}. We leave this as our future work.  Figure~\ref{fig:tl_qoe}b shows the error
rate when applying TL to a QoE model from a different group to the LEG group. Like the previous experiment, we see the accuracy improvement
reaches a plateau when using 18 webpages.


\subsection{Evaluation of Continuous Learning}
To mimic the impact of changing web workloads on a deployed QoE or FPS predictor, we  train an \emph{initial} predictor
on 50\% of the training samples and test the trained predictor on the remaining webpages using cross-validation. To
isolate the impact of TL, the initial
 models in this evaluation are learned using data collected from the target environment.

\cparagraph{Detect ageing models.} Our first experiment aims to evaluate \SystemName's ability in using CP to detect an
ageing QoE predictor due to workload changes. We do not apply CP to the FPS predictor because the ground-truth can be
directly measured. We are interested in knowing how often our CP function (see Section~\ref{sec:agining}) successfully
detects when a predicted QoE target has an error of more than 5\%. Our CP scheme successfully catches 96.4\% of the
inputs where the QoE predictor gives a wrong prediction under our criterion. Our scheme also has a low false positive
(i.e., when the CP model thinks the QoE predictor is wrong but it is not) rate of 5\%.

\cparagraph{Model update.} We can use user feedback (for QoE predictions) or automated profiling information (for FPS
predictions) on the first few  mispredicted webpages flagged by \SystemName to update an existing model. We found that
\SystemName updated using five mispredicted webpages delivers on average 98\% (for QoE predictions), and 97\% (for FPS
predictions) of the performance given by a model trained using the entire dataset on the target platform. This
translates into an improvement of over 23.4\% for the initial predictor in this experimental setting. Because profiling
only needs to be performed on incorrectly predicted inputs, the model retraining process is fast, taking less than 2
minutes on a XiaoMi 9 phone; in comparison, profiling on the entire training dataset would take hours.

In practice, one would first use TL to tune the baseline predictors during the first installation. Then, the CF scheme
can be used to update the installed models. This experiment shows that \SystemName is highly effective in detecting and
updating ageing models without incurring significant overhead.

\subsection{Model Analysis}
\label{sec:analysis}

\begin{figure}[!t]
	\centering
	\subfloat[][\# Training hidden layers]{\includegraphics[width=0.23\textwidth]{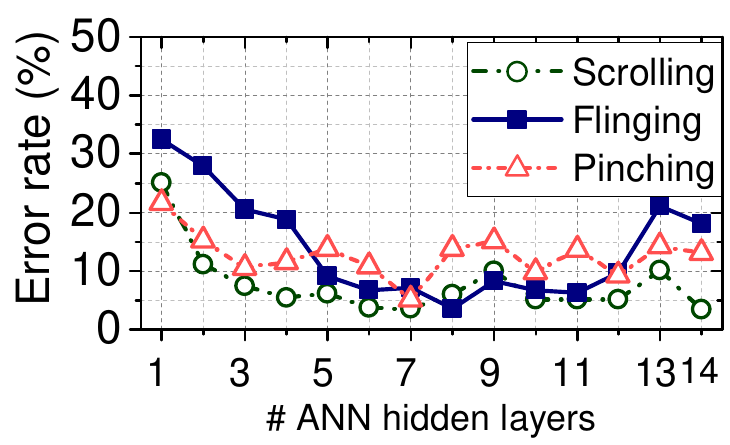}}
    \subfloat[][Perf. w.r.t. \#webpages]{\includegraphics[width=0.23\textwidth]{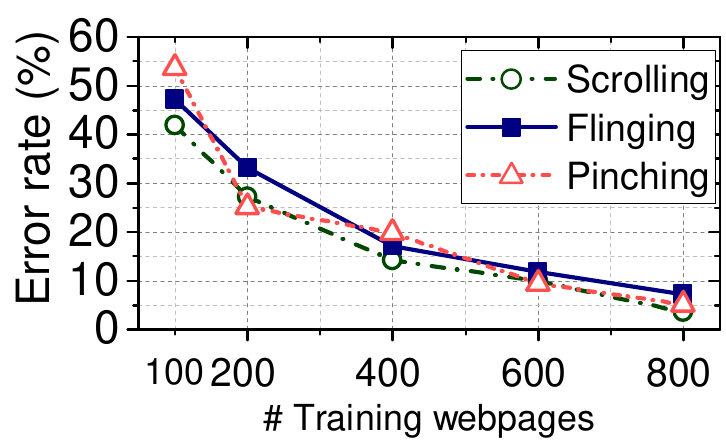}}
    \caption{Impact of the number of hidden neural layers (a) and training webpages for our ANN-based FPS predictors (b).}
    \label{fig:num_layers_samples}
\end{figure}

\begin{figure*}[t]
    \centering
    \subfloat[][Xiaomi 9]{\includegraphics[width=0.45\textwidth]{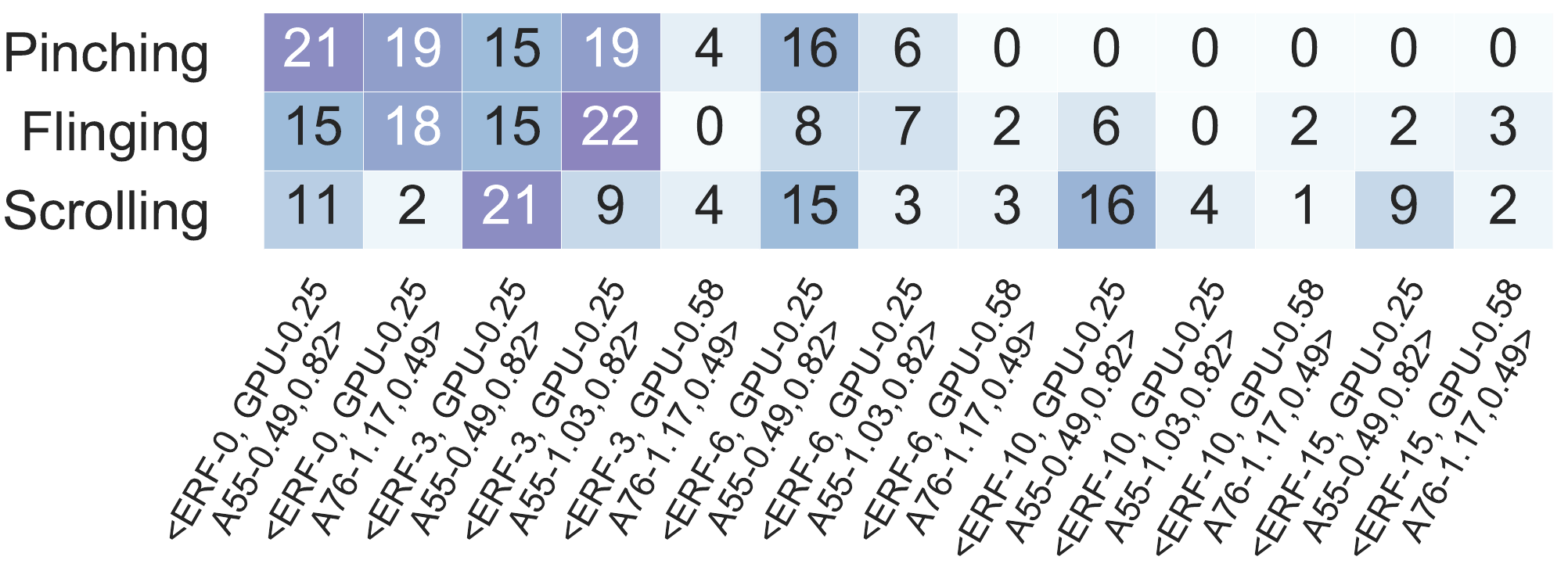}}
    \subfloat[][Google Pixel 2]{\includegraphics[width=0.42\textwidth]{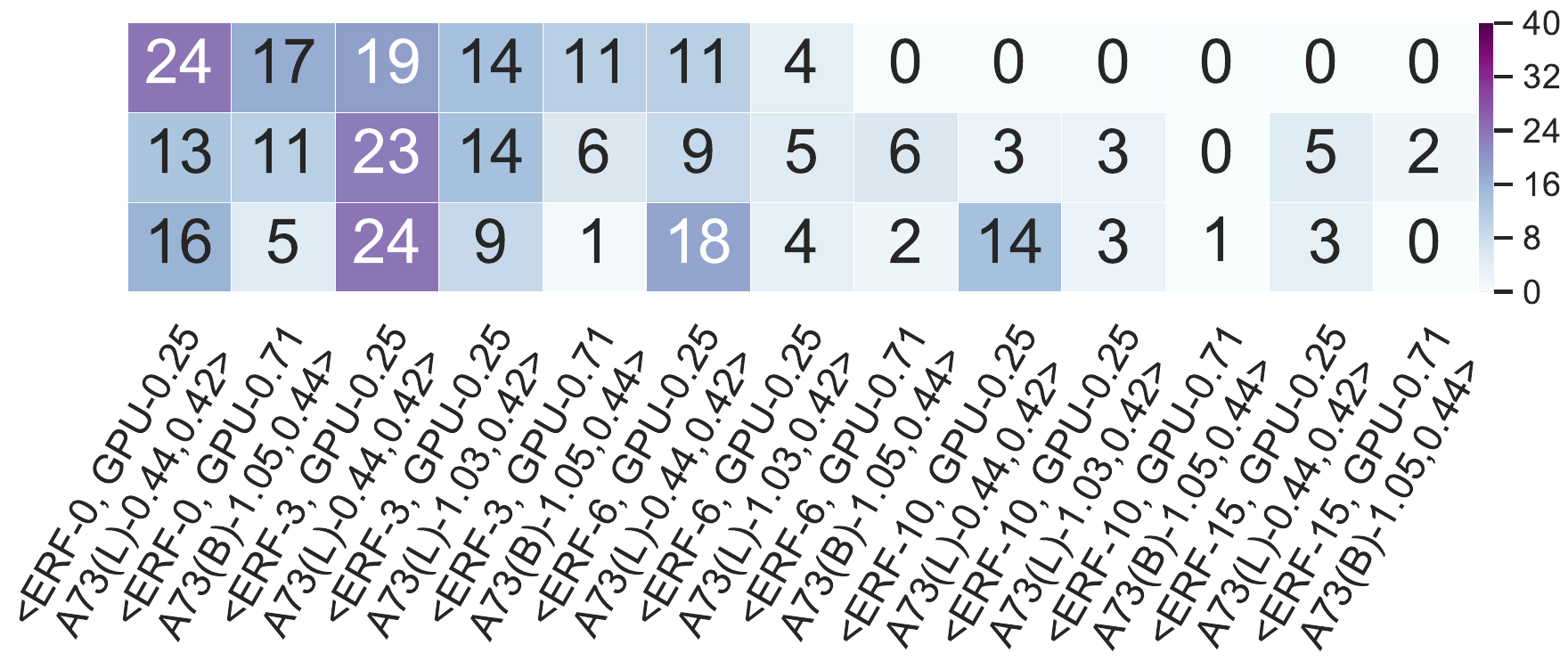}}
    \\
    \subfloat[][Huawei P9]{\includegraphics[width=0.45\textwidth]{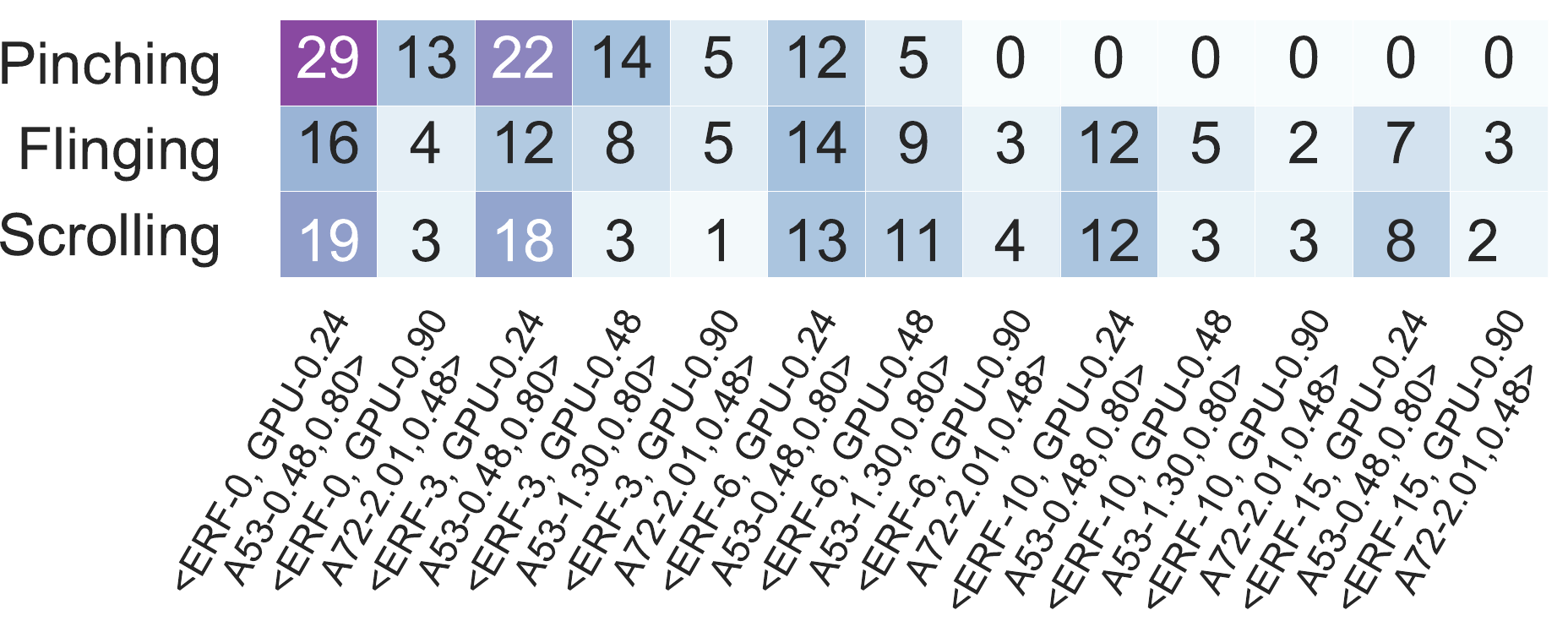}}
        \subfloat[][Odroid Xu3]{\includegraphics[width=0.45\textwidth]{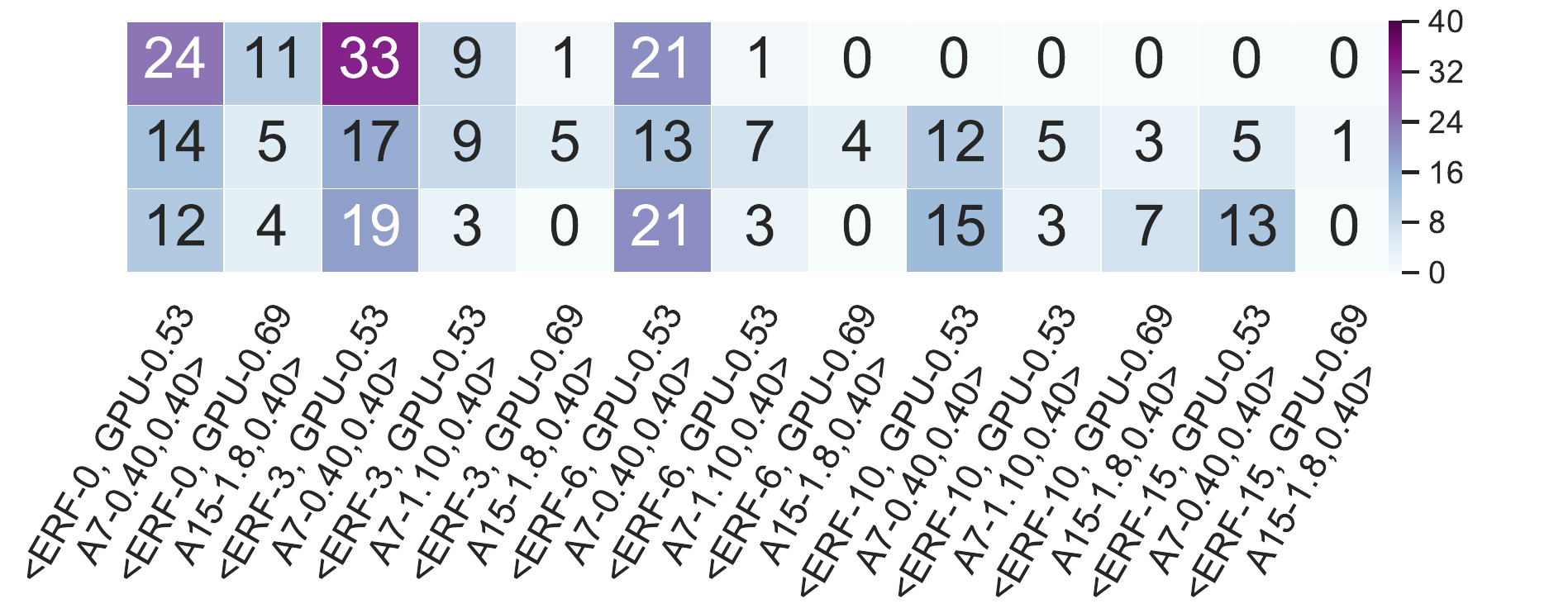}}
    \caption{Distributions of optimal processor settings.}
    \label{fig:tl_dist}
\end{figure*}

\subsubsection{Impact of neural layers and training samples} Figure~\ref{fig:num_layers_samples}a gives the error rate when an ANN-based
FPS predictor is constructed with different numbers of hidden layers. To isolate the impact of TL, we first train the model using 800
webpages and then test the trained model on another 200 webpages. Using 7, 14 and 8 hidden layers give the best performance for scrolling,
flinging and pinching respectively. We choose to use a unified model structure with 7 hidden layers as it requires fewer training examples
and the performance is not far from the optimal settings.  Looking at Figure~\ref{fig:num_layers_samples}b, we see a steady decline in
error rates when using more examples to train the baseline predictors. This is not surprising, as the performance of a predictive model
generally improves as the number of training samples increases. Since a baseline model only needs to be trained once, this is a one-off
cost.



\subsubsection{Processor configuration distributions} Figure~\ref{fig:tl_dist} shows the distribution of the most optimal processor
settings. Here, we use the notation $<$ ERF - event response frequency, GPU-freq, rendering CPU core - rendering CPU
core freq, other CPU core freq$>$ to denote a processing configuration. For example, $<$ERF-10, GPU-0.48, A53-1.3,
0.8$>$ means that we response to 1 out of every 10 input events of the same type, the painting process running on the
GPU at 480MHz, and the render process running on the little A53 core at 1.3 GHz while the big core operates at 800MHz.
Although some of the configurations are being optimal more frequently than others, the distribution varies across event
types and hardware platforms. This diagram reinforces the need for an adaptive scheme. \SystemName is designed to offer
such a capability.

\subsubsection{Alternative modeling techniques} 
\begin{figure}
    \begin{minipage}[t]{0.46\linewidth}
    \includegraphics[width=1\textwidth]{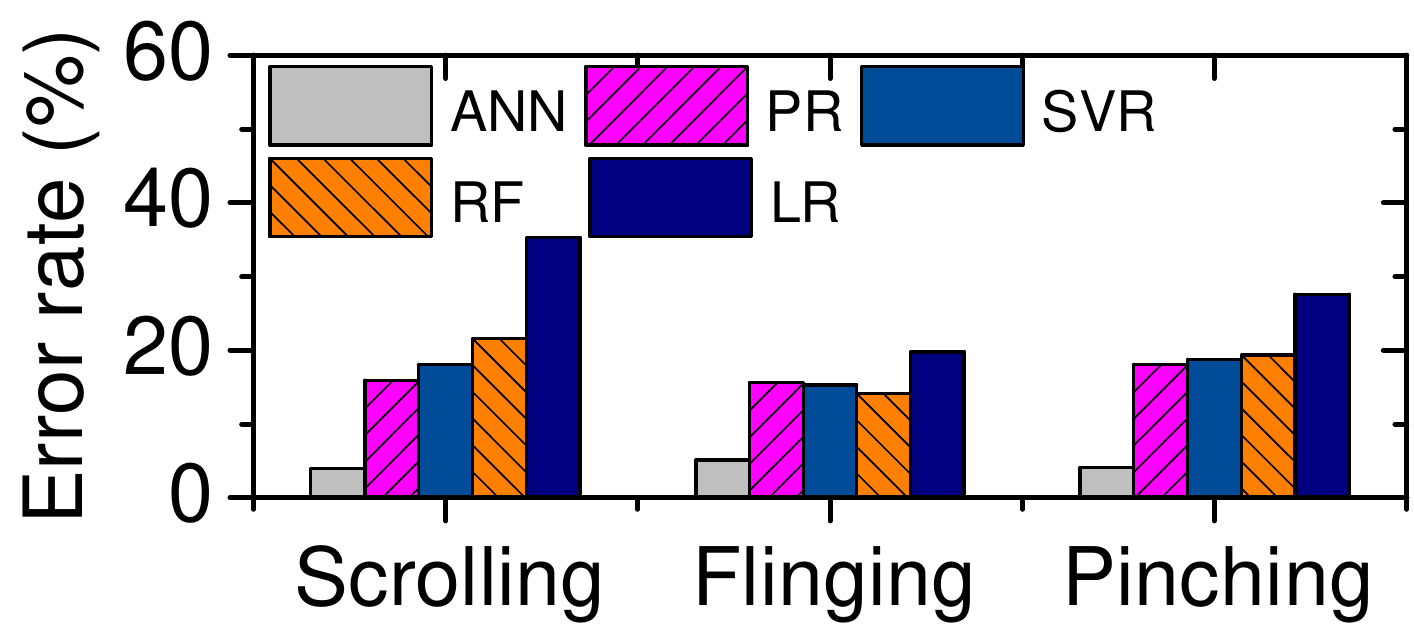}
    \caption{Comparing our ANN-based predictor with other modeling techniques.}
    \label{fig:alternative}
  \end{minipage}
  \hfill
  \begin{minipage}[t]{0.44\linewidth}
        \includegraphics[width=1.05\textwidth]{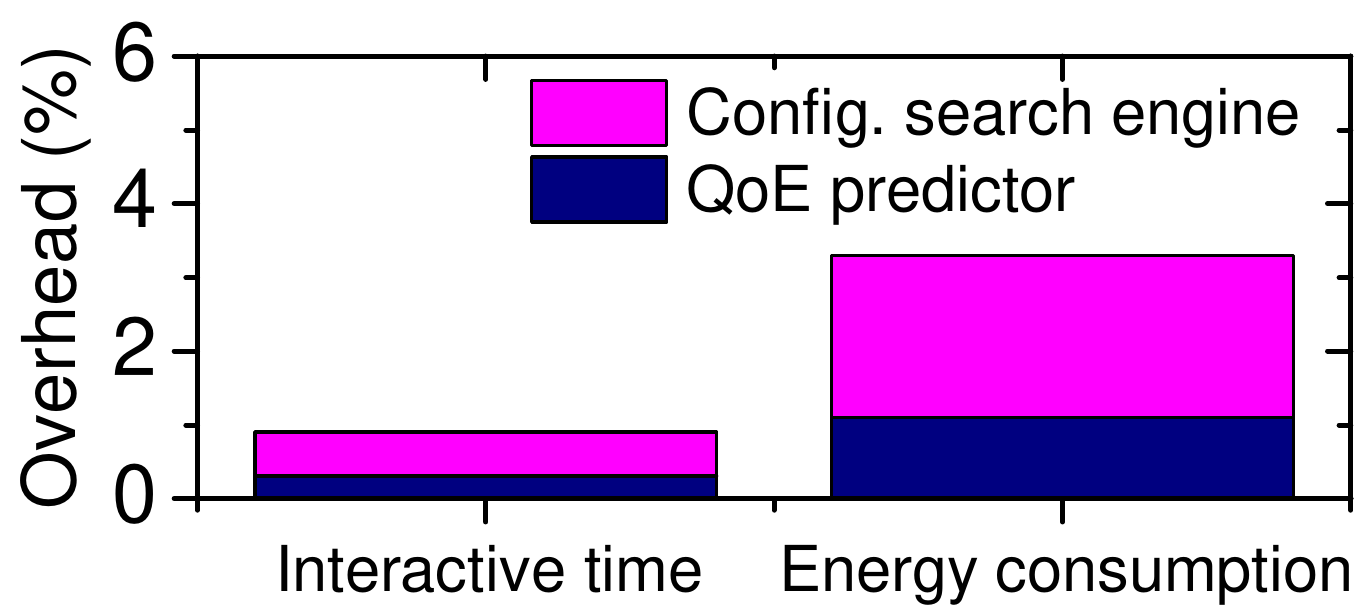}
    		\caption{Breakdown of runtime overhead.}
    \label{fig:overhead}
  \end{minipage}
\end{figure}

Figure~\ref{fig:alternative} compares our ANN-based FPS predictor
against four alternative regression methods used in prior relevant works: Linear Regression (LR), Polynomial Regression (PR), Support
Vector Regression (SVR), and Random Forest (RF). All the alternative techniques were trained and evaluated by using the same method and
training data as our models. Our approach achieves the lowest error rate and enables us to employ transfer learning.

\subsubsection{Overhead breakdown}
Figure~\ref{fig:overhead} gives a breakdown of the runtime overhead of \SystemName (which was already included in our experimental
results). \SystemName introduces little overhead to the end to end turnaround time and energy consumption, less than 1\% and 4\%
respectively.

\subsection{Discussions and Future Work}
\cparagraph{Multi-tasking environment.} \SystemName can be extended to a multi-tasking environment for optimizing the front-running
application. On mobile systems, background workloads are typically put into a sleeping or closed status, and thus not require a quick
response at the background. \SystemName can also be integrated with an interference-aware scheduler like~\cite{8366936} to minimize the
impact on concurrently running workloads.

\cparagraph{Display optimization.} Our experimental results already include energy consumption of the screen, but we do not optimize the
display setting. Since the display setting does not affect the processing latency, \SystemName can be easily integrated with a display
optimization scheme like ~\cite{shye2009into} and \cite{6148235}.

\cparagraph{Dynamic content.} Our work does not consider network latency as most of the web content would already be downloaded before a
user interaction commences. However, it is possible that user interaction will trigger new network activities. Methods on latency-aware
optimizations for page loading~\cite{Ren:2018:PNW:3281411.3281422} or dynamic content~\cite{mehrara2011dynamic} are thus complementary to
\SystemName.

\cparagraph{Apply to other applications.} \SystemName can be directly applied to WebView-based applications without modification to the
application code. From Android 4.4, WebView is based on the Chromium rendering architecture on which \SystemName is tested.

\section{Related Work}
Our work is broadly related to the literature in four areas:

\cparagraph{Mobile web workload optimization.} Prior work has focused on the initial page loading phase through e.g.,
dynamic frequency scaling~\cite{ren2017optimise,Ren:2018:PNW:3281411.3281422,8366936}, accumulating
traffics~\cite{6848020, li2016automated} and parallel downloading ~\cite{sehati2017energy}. \SystemName targets the
later user interaction phase. It is closely related to event-based power management for mobile web
browsing~\cite{zhu2015event,Peters:2018:PWB:3205289.3205293, ml2019}. However, these previous methods have three
drawbacks: (1) by assuming a fixed response deadline, (2) have intensive overhead for targeting new hardware and user,
and (3) cannot examine whether a decision model still fits. eBrowser~\cite{xu2018ebrowser} uses image entropy to
characterize the web content, but it requires all web contents to be rendered ahead of time, introducing significant
start up delays and could waste computation cycles. \SystemName is designed to address these limits, offering a better
and practical way to target a wider range of computing environments.

\cparagraph{QoE modeling.} Prior research models user experience through usability studies~\cite{rogers2007s},
contextual inquiries~\cite{ferreira2014contextual} or data logging~\cite{Zuniga:2019:THQ:3308558.3313428}, by
considering generic metrics like power consumption, response time and network latency. Unlike these works, \SystemName
is a content-aware QoE estimation scheme by directly modeling the impact of web workloads on QoE.

\cparagraph{Energy optimization.} Other relevant works include optimizations for the display~\cite{he2015optimizing}
and radio~\cite{zhao2014energy}, dynamic content caching~\cite{Zare:2016:HTS:2964284.2967292} or
prefetching~\cite{Bui:2015:RET:2789168.2790103}, optimizations for JavaScript~\cite{mehrara2011dynamic}, and
multi-event scheduling~\cite{pes}. As pointed out in~\cite{Bui:2015:RET:2789168.2790103}, mobile web browsing requires
novel techniques across the computing stack; \SystemName thus benefits from techniques from different computing layers.

\cparagraph{Machine learning for systems optimizations.} Machine learning has been used to model power
consumption~\cite{Hoque:2015:MPD:2856149.2840723}, task scheduling~\cite{mlcpieee,taylor2017adaptive} of mobile systems and program tuning
in
general~\cite{wang2014integrating,Tournavitis:2009:THA:1542476.1542496,Wang:2009:MPM:1504176.1504189,wang2010partitioning,grewe2013portable,wang2013using,DBLP:journals/taco/WangGO14,
ogilvie2014fast,wen2014smart,cummins2017end,ogilvie2017minimizing,spmv,ipdpsz18,ijpp18,tecs19,grewe2011workload,emani2013smart,grewe2013opencl,adaptivedl,zhang2020optimizing}.
Our work tackles an outstanding problem of porting a model to a new computing environment. Transfer learning was recently used for wireless
sensing~\cite{crosssense} through randomly chosen samples. \SystemName improves ~\cite{crosssense} by carefully choosing representative
tracing examples for transfer learning. Conformal prediction was used for malware classification~\cite{jordaney2017transcend}, but not the
regression problem addressed by \SystemName. We note that the novelty of \SystemName is a new way of combining statistical learning and
techniques, rather than improving the learning algorithm itself.

\vspace{-1mm}
\section{Conclusions}
\vspace{-2mm} This paper has presented \SystemName, a novel energy optimization scheme for interactive mobile web browsing. Unlike prior
work, \SystemName models how the web content and interactive speed affects the QoE. To develop a practical solution, \SystemName employs
transfer learning and conformal predictions to automatically adapt an existing policy to the changes of users, hardware platforms or web
workloads. We apply \SystemName to Chromium and evaluate it on four mobile systems across 1,000 webpages and 30 users. Experimental results
show that \SystemName consistently outperforms existing web-optimizers, and has less overhead when targeting a new user or device.

\section*{Acknowledgements}
This work was supported in part by the NSF China under grant agreements 61902229, 61872294, 61877038, 61602290 and 61877037; and the
Natural Science Basic Research Program of Shaanxi Province under Grant No. 2019JQ-271. For any correspondence, please contact Zheng Wang
(E-mail: z.wang5@leeds.ac.uk).

\balance{}
{\small
\bibliographystyle{IEEEtran}
\bibliography{refs,zheng2}
}

\end{document}